\documentclass[11pt]{article}

\usepackage[top=1in,bottom=1in,left=1in,right=1in,a4paper]{geometry}
\usepackage{authblk}

\usepackage{hyperref}

\usepackage{amsfonts,amsthm,amssymb,amsmath}
\usepackage{color,graphicx,mathrsfs}
\usepackage{url}

\usepackage{algorithmic}
\usepackage{imakeidx}
\usepackage{multirow}
\usepackage{float}
\usepackage{tikz}
\usepackage{pgfplots}
\usetikzlibrary{calc,shapes,positioning,arrows.meta}
\usepackage{caption,subcaption}

\usepackage[numbers]{natbib}
\newcommand{\prob}{\mathbb{P}} 
\newcommand{\Ex}{\mathbb{E}} 

\newcommand{\tx}{{\texttt{Tx}}} 
\newcommand{\sdag}{{\mathcal{G}}} 
\newcommand{\lev}{{\mathcal{S}}} 
\newcommand{\leaf}{{\mathcal{T}}} 
\newcommand{\confirm}{{\mathcal{C}}} 

\newcommand{\gen}{{\texttt{O}}} 
\newcommand{\blk}{{\texttt{B}}} 
\newcommand{\idp}{{{\texttt{id}}_{\textrm{prev}}}} 
\newcommand{\idt}{{{\texttt{id}}_{\textrm{tip}}}} 
\newcommand{\idm}{{{\texttt{id}}_{\textrm{ms}}}} 
\newcommand{\peer}{{\texttt{peer}}} 
\newcommand{\pow}{{\texttt{nonce}}} 
\newcommand{\mes}{{\texttt{message}}} 

\newcommand{\trgt}{{d}} 
\newcommand{\height}{\eta} 

\newcommand{\utxo}{{\mathcal{U}}} 
\newcommand{\ledger}{{\mathcal{L}}} 

\title{
  Consensus Mechanism Design \\ based on Structured Directed Acyclic Graphs
}

\author[*]{Jiahao He}
\author[*]{Guangju Wang}
\author[*]{Guangyuan Zhang}
\author[$\dagger$]{Jiheng Zhang}

\affil[*$\dagger$]{DMAC Lab}
\affil[*$\dagger$]{Department of Industrial Engineering and Decision Analytics}
\affil[$\dagger$]{Department of Mathematics}
\affil[*$\dagger$]{The Hong Kong University of Science and Technology}

\date{\today}

\begin{document}
\maketitle

\begin{abstract}
  We introduce a structure for the directed acyclic graph (DAG) and a mechanism design based on that structure so that peers can reach consensus at large scale based on proof of work (PoW). 
  We also design a mempool transaction assignment method based on the DAG structure to render negligible the probability that a transaction being processed by more than one miners. 
  The result is a significant scale-up of the capacity without sacrificing security and decentralization. 
\end{abstract}

\emph{
  Key words: 
  consensus;
  directed acyclic graph;
  proof of work; 
  transaction assignment.
}


\section{Introduction}

Blockchain technology is fundamentally a consensus mechanism design based on a chain structure of storing information in a distributed manner over a peer-to-peer network.
Taking Bitcoin \cite{nakamoto2008bitcoin} for example, multiple transactions are grouped together and stored in a block that is less than 1 megabyte in size, and all blocks are connected in a chain structure.
Although this may not be the most efficient way of storing information, with the help of cryptographic tools such as hash functions and public/private keys for integrity check and authentication, Nakamoto consensus \cite{nakamoto2008bitcoin} proposed based on the chain structure is able to achieve both security and decentralization.
Blockchain technology has achieved great success as exemplified by Bitcoin \cite{nakamoto2008bitcoin} and Ethereum \cite{wood2014ethereum}.
They have proved that a decentralized and secure public ledger system is not only possible, but also has a great impact in our financial system and gives rise to a whole new ecosystem with extendable services and applications.

In Bitcoin, all participants compete in solving cryptographic puzzles by tuning a nonce so that the hash result exhibits a certain required pattern known as proof of work (PoW).
The puzzles are so difficult that in a given period of time, long enough for a block of a given size to propagate over the network, it is likely only one miner can form such a block.
%
Miners, in fact those who will create future blocks, can choose not to agree with the information in a block simply by forking from a position preceding that block. 
Based on the honest majority assumption, all honest miners will eventually be able to collectively agree upon a chain of valid blocks, with invalid blocks on the forks. 
So all blocks ever created form a tree.
The longest chain in the tree, referred to as the Nakamoto chain, is collectively created by all honest miners using PoW.
It is the Nakamoto chain that confirms all the historical records.
A rigorous formulation of consensus has been proposed by \cite{garay2015bitcoin} together with a probabilistic model that formally proves it.
With its rapid adoption over the last decade, blockchain technology has now come face to face with a serious bottleneck, the extremely limited capacity, i.e., small number of transactions per second (TPS). 
Removing this bottleneck will be a significant breakthrough in advancing the blockchain technology, and open the possibility for a wide range of applications.
However, scaling up the capacity should not be at any compromise of security and decentralization.
 
Within the confines of the chain structure, we can try to increase the block size, or equivalently to decrease the time interval between blocks, or use a combination there of to increase the TPS.
However, since larger blocks require more time to propagate over a peer-to-peer network, such an attempt would increase the occurrence of forked blocks, even in the absence of malicious miners. 
In order to increase the capacity by an order of magnitude, it is inevitable that we explore beyond the chain structure. 
This naturally leads to the idea of expanding the chain of blocks to a directed acyclic graph (DAG) of blocks to allow \emph{parallelism}. 
The idea has been implemented by IOTA \cite{popov2016tangle}.
In IOTA, everyone has the right to form a small block as no PoW is required. The validity of a block is justified whenever someone is willing to append a new block to it either directly or indirectly. 
This requires performing the computationally expensive task of updating weights to calculate the number of blocks appended to each block.
The weights update approaches such as Ghost \cite{sompolinsky2015secure} suffer from what is known as the  ``balancing attack'' \cite{bagaria2018deconstructing}.
Meanwhile, capacity is not the only issue in current blockchain systems.
With the ever-growing hashing power, cryptographic puzzles are becoming increasingly difficult. 
Miners with certain hashing power would be able to mine a certain number of blocks and collect the corresponding reward in expectation.
However, the huge fluctuations of the actual reward around the expectation force most miners to join mining pools in order to smooth their income, leading to the concentration of hashing power within a few big mining pools. 
Another issue is the high latency since it takes a long time to confirm a transaction. 
In addition, because they are self-interested, miners would try to pack transactions with high fees into their blocks to maximize their reward. 
Thus, transactions with little to no transaction fees hardly have a chance to be processed. 

Among the different ways to achieve security and decentralization, PoW and Nakamoto consensus are two fundamental tools that have been proven to be effective.
In this paper, we further explore the idea of expanding the chain to a DAG, while still keeping using these two fundamental tools. 
Without sacrificing security and decentralization whatsoever, the objectives are to
\begin{itemize}\setlength\itemsep{0pt}
  \item scale up the capacity;
  \item shorten the latency;
  \item deconcentrate the mining power;
  \item increase the probability that transactions with small fees are processed.
\end{itemize}
%
%
Our key idea is to embed a Nakamoto chain in a DAG by designing a structure that is strongly connected and incorporates miner information.
In this way, the security of our design is guaranteed by the security of the proven Nakamoto consensus, eliminating the need to update weights. 
Based on the guaranteed security, we will take advantage of the strong connectivity in our DAG structure to increase the throughput and shorten the latency.
A DAG also provides an ordering of transactions so that all honest peers (miners)
\footnote{We use the terms``miner'' and ``peer'' almost interchangeably. The only subtle difference is that a miner must solve  cryptographic puzzles to create blocks, while a peer does not. Like a miner, a peer may propagate blocks to a peer-to-peer network.}
will be able to build the same public ledger once they have reached a consensus on the DAG.
We will also explore the rich possibility of using our designed structure in DAGs to realize various improvements in line with the above objectives.

Our design breaks a large block into multiple smaller ones. 
In fact, for simplicity, we put only one transaction in each block. 
This does not only lead to a smaller block size, but also much easier cryptographic puzzles.
This design enables miners to broadcast transactions (stored in small blocks)  continuously over time, instead of waiting to broadcast a large batch of transactions every once in a while.
As mentioned earlier that a major bottleneck for blockchains is their linear structure which forbids parallelism and the tradeoff between block size and synchronization time.
Using small blocks allows fast peer-to-peer propagation and parallelism, thus significantly improving the throughput.
However, a well-designed structure in the DAG is needed together with a consensus mechanism for security and robustness.
This leads to our idea of interspersing relatively more difficult blocks, referred to as \emph{milestones}, in the DAG for security purpose. 
Our DAG starts from the genesis block, which contains a set of trusted setup information. 
The workflow is the same for creating every block, regardless of whether it will be a milestone or a regular block.
When preparing a new block, every miner must choose a transaction that is valid to the best of his latest knowledge, and specify three pointers.
The first pointer points to the miner's previous block, or the genesis if the miner does not have any previous block. With this requirement, each miner will have a \emph{peer chain} representing the state of that miner, which enables the possibility of incorporating information such as the miner's ``identity'' and hashing power. 
The second pointer points to the previous milestone, or the genesis if there is no previous milestone, following the traditional longest chain principle in the case where there is a fork.
This pointer is required because the miner does not know whether or not the new block will turn out to be a milestone. If the new block does turn out to be a milestone, we will need all of the milestones to be connected to form a Nakamoto chain. 
The last pointer must point to another miner's recent regular block to enhance connectivity among peer chains. 
Note that pointing to too old a block contribute little to connectivity and thus is undesirable.
The reward scheme in our protocol incentivizes miners to connect to recent blocks.
Our DAG structure is illustrated in Figure~\ref{fig:dag}, where peer chains and the Nakamoto chain are highlighted. 

After preparing a block, the miner will simply hash this block by tuning the nonce. If the hashing result exhibits a certain required pattern, e.g.\ the 10 leading bits are all 0, then it is a valid block. If by luck, the hashing result exhibits a more difficult pattern, e.g.\ the 15 leading bits are all 0, then it is a milestone block.
A block, regardless of its type, serves as a transaction container.  
The milestone blocks form a Nakamoto chain, which has the additional role of assisting peers in reaching a consensus.
Note that whether or not a block is a milestone is random and only revealed when the mining process is completed. 
No miner can devote his hashing power exclusively for milestone blocks since the workflow for creating any block is the same and the hash function has three desirable properties --- collision-free, hiding and puzzle-friendly \cite{narayanan2016bitcoin}.

Such a design enables scaling up the capacity.
Moreover, the milestone chain is completely decentralized and enables reaching a consensus in the same way that the Nakamoto chain does.
The latency is reduced since a new block is soon confirmed by a milestone as analyzed in Section~\ref{sec:analysis}.
By ``confirm'' we mean that there is a path from a milestone block to this new block by following a sequence of the above-mentioned pointers.
A smaller block yields a smaller reward but it also requires a smaller hashing effort.
This greatly reduces the variability in mining rewards, thus eliminating miners' need to rely on mining pools to smooth their income.

As the capacity is scaled up, there is a higher chance that a transaction will be processed by more than one miner during a short time interval.
Although the conflict can be resolved by consensus, it is a waste of capacity --- something we are trying so hard to increase --- to have multiple blocks containing the same transaction.
Thus we need a mechanism to avoid this kind of collision. 
The main idea is to divide all of the transactions in the mempool among the miners according to their hashing power.
This requires knowledge of their identity and hashing power, which can be inferred from the peer chain design. 
A miner can only process a transaction if the hashing result of its most recent block concatenated with the transaction satisfies a certain pattern, e.g.\ the leading several bits are 0. 
Miners are always allowed to create ``empty'' blocks containing no transaction only earn a reward for creating blocks but no transaction fee.
By carefully choosing the parameters, this method effectively reduces the waste of capacity without sacrificing too much the waiting time of transaction in the mempool, as shown in our analysis in Section~\ref{sec:analysis}.


Having introduced the above ideas of scaling up the capacity, we will now discuss the physical ceiling of capacity. 
One constraint comes from network synchronization. 
For a public ledger system with a specified TPS, data will be accumulated at a corresponding speed. 
One way or another, this amount of data needs to be synchronized among all peers over the current Internet infrastructure.
In fact, the network synchronization affects both throughput and latency.
Another constraint comes from computing. For example, what is the fastest speed of writing and reading a database (e.g., LevelDB). We also find that the \texttt{secp256k1 ECDSA} key verification is computationally expensive and could potentially be a bottleneck as TPS increases. 
Our mechanism design can help a system to approach the ceiling.
Our design does not require strong information synchronization as in some current systems. 
Unlike systems requiring frequent weights updates, our design does not require expensive computing.
Putting only one transaction in a  block may seem inefficient, but the amount of overhead created is in fact quite small due to the registration and redemption design introduced in Section~\ref{sec:DAG-to-Ledger}, and so the design is worthwhile considering the advantages it brings. 
How close our design is to the ceiling can only be verified through an open test net, which is under development at the time of writing. 
We hope to add running results of our open test net in future version updates.

Our mechanism design can be viewed as an extension of blockchain technology, such as Bitcoin \cite{nakamoto2008bitcoin} and Ethereum \cite{wood2014ethereum}, and unstructured DAGs, such as IOTA \cite{popov2016tangle}.
Recently, building upon the idea of DAG, Conflux \cite{li2018scaling}solves the waste of discarding the fork blocks by selecting the ``pivot chain'' with the highest ``weight'' whose consensus is guaranteed by Ghost \cite{sompolinsky2015secure}.
In contrast, we eliminate the need to compute weights and use Nakamoto consensus instead.
Algorand \cite{chen2016algorand} provides a new scaling consensus based on Byzantine Agreement, and Thunderella \cite{pass2018thunderella} offers a new consensus mechanism that is robust in the worst case and enables fast transaction confirmation in the optimistic case. 
These topics are beyond the scope of this paper.
Many problems exist in the emerging field of blockchain and cryptocurrency. 
For example, as TPS increases to the thousands, a dozen terabytes of data will easily accumulate each year. The current reward scheme only incentivizes people to provide mining service to a public ledger. 
How do we design an economic model that incentivizes people to provide storage and the required bandwidth? 
We will leave this kind of problem to future research.

\section{A Structured DAG}
\label{sec:dag}

Among many applications, a graph can describe a data structure, with each vertex representing a basic unit of information, and each edge the relationship between two units.
In this section, we describe in detail our proposed structure in a DAG. 
The objective of this structure is to enable miners across the network to reach a consensus on a set of data, so that they can build and maintain a public ledger in a distributed fashion that all honest peers agrees on. 
We present the DAG structure in a general setting without too much dependency on public ledgers, since it can potentially have other applications. 


We first describe the basic element, namely a \emph{block}, in the DAG.
As a basic unit of information, a block $\blk$ should contain a \mes{}, which is the essential data such as transactions for cryptocurrency applications. 
It should also contain some additional information, namely the \emph{block header}, for integrity checking and positioning in the DAG.
Assume there is a random oracle $H$ which maps a string of arbitrary length to a unique identity.
In actual implementation, we would use a reasonably good cryptographic hash function, e.g. \texttt{SHA-256}, as the random oracle.
We apply the random oracle to the concatenation of all parts of a block, symbolically denoted by $H(\blk)$, to obtain the block's identity. 
To check a block's integrity, we just need to check whether the block and its identity match under the random oracle.  
Formally, a block 
\begin{equation}
  \label{eq:block}
  \blk = (\idp, \idm, \idt, \peer, \pow, \mes),
\end{equation}
where ($\idp,\idm$, $\idt$) are the unique identities of some other blocks in the DAG, \peer{} is the miner who creates on this block, and \pow{} is the solution to the cryptographic puzzle.
Before giving further explaining the components of a block, we define \textit{the genesis} $\gen$ to be a special block containing certain trusted setup information.

\paragraph{\bf Proof of Work.}
We require a proof of work (PoW) for each block in order to decide if a miner has the right to form the block. 
The miner should tune the \pow{} until the hash of the block $H(\blk)$ exhibits a certain pattern.
Suppose the hash result $H(\blk)$ is a string of zero-one bits and denote by $.H(\blk)$ the number in $[0,1]$ to which this string of bits converts following the convention in \cite{chen2016algorand}. 
Since $H$ is a random oracle, $.H(\blk)$ can be modeled as a random variable uniformly distributed on the interval [0,1] for any block \textit{a priori}.
In order to ``prove work'', a miner needs to vary the \pow{} until $.H(\blk)<\trgt$, where $\trgt$ is called the difficulty.
Among all the blocks, we allow a portion of them to be of a special type, called milestones.
A block is a milestone if $.H(\blk))<p\trgt$, where $p$ is the probability that a block is a milestone.
Note that a miner has to specify all parts of a block including the three pointers, the main message and the nonce before working on it (computing the hash).
The type of block is revealed only after the peer has worked on it. In other words, a peer does not know whether he is working toward a milestone or a regular block.
A miner may continue working on the block after obtaining a hash $.H(\blk)\in[p\trgt,\trgt)$ with the intention of making it a milestone. 
However, devoting a miner's hashing power to the creation of milestone blocks will not change the expected number of milestone blocks he is able to create. 
Since a regular block also yields a mining reward, albeit a smaller reward than that offered by a milestone, and involves a transaction fee, exclusively mining milestone blocks is not economically beneficial for peers.

To describe the structure we want to design in a DAG, we now explain how the three pointers $(\idp, \idm, \idt)$ position the block in the DAG and the intuitions behind our design.

\paragraph{\bf Peer Chain.}
The first pointer $\idp$ points to the most recent block created by the same miner or the genesis if the miner has not mined any blocks before. 
By this mechanism, blocks mined by the same miner are organized into a chain, namely the \emph{peer chain}.
The head of the chain, defined to be the most recent block, of the chain can also be interpreted as the state of the miner. 
A peer chain is designed to provide not only a clear structure in the DAG, but also valuable information including the miner's mining history, from which we can estimate the miner's hashing power. 
In Section~\ref{sec:DAG-to-Ledger}, we will introduce a transaction scheduling scheme that utilizes the miner state and estimated hashing power to reduce the probability that a transaction is processed by multiple miners.
The peer chain could also potentially generalize our design to credit-based applications.
Formally, we require
\begin{equation}
  \label{eq:idp}
  \blk.\idp = H(\blk')
  \implies 
  \{\blk'.\peer = \blk.\peer\} 
  \textrm{ or } 
  \{\blk' = \gen\}.
\end{equation}
Note that this mathematical requirement cannot prevent a miner from forking his own peer chain (e.g., not appending to the miner's most recent block) nor attacking (e.g., forking another miner's peer chain).
As explained later in the reward scheme in Section~\ref{sec:DAG-to-Ledger}, a miner not appending to his most recent block only cause less mining reward and waste of his hashing power.
And a miner mining on another miner's peer chain might be wasting a small amount of the system capacity at the cost of his own hashing power without any benefit to anyone.
None of the above actions would affect the consensus.
In fact, incentivizing a miner to append a newly mined block to his most recent block lessens the effect of ``lazy connecting'', i.e.\ not appending to recent blocks, an issue raised in the implementation of IOTA \cite{popov2016tangle}.
The pointer $\blk.\idp$ specified in \eqref{eq:idp} is essentially a directed edge from $\blk$ to $\blk'$ if we view each block as a vertex in a graph. 
Any directed edge plays a confirmation role in that block $\blk$ confirms $\blk'$, which confirms its previous blocks via its three pointers. 

\paragraph{\bf Connectivity.}
To create the connectivity among different peer chains, we use $\idt$ to point to a regular block of another miner. The genesis $\gen$ is set as the default when no such block exists. Formally, 
\begin{equation}
  \label{eq:idt}
  \blk.\idt = H(\blk_t) 
  \implies 
  \{\blk_t.\peer \neq \blk.\peer,\ .H(\blk_t)\in[p\trgt, \trgt)\}
  \textrm{ or } 
  \{\blk_t = \gen\}.
\end{equation}
Again, the directed edge from block $\blk$ to $\blk_t$ plays a role of confirming $\blk_t$ and all of the blocks that $\blk_t$ confirms.
See \eqref{eq:confirm} for a mathematical definition of confirming previous blocks.
Intuitively, a stronger connection leads to faster confirmations, thus we could have multiple such pointers to further enhance the connectivity. 
However, these pointers have to be synchronized, verified and stored by all peers and therefore having more pointers will incur a higher overhead cost.
Our analysis in Section~\ref{sec:latency} shows that having one such pointer is enough to ensure reasonable confirmation latency.

\paragraph{\bf Embedding a milestone chain.}
The pointer $\idm$ points to a milestone block $\blk_m$ or the genesis $\gen{}$, i.e.
\begin{equation}
  \label{eq:idm}
  \blk.\idm = H(\blk_m) 
  \implies 
  \{.H(\blk_m)<p\trgt\}
  \textrm{ or } 
  \{\blk_m = \gen\}.
\end{equation}
Our milestone chain works the same ways as the Bitcoin blockchain. 
The difference is that each blocks on our milestone chain is smaller and confirms some other regular blocks in a structured way. 
A major barrier to scaling up the capacity of a blockchain system is the slow synchronization of large blocks in a peer-to-peer network.
Our milestone chain, consists of much smaller blocks, is designed as a bridge between high throughput and stable synchronization. 
Each milestone, despite being small, confirms a relatively large number of other blocks as defined in \eqref{eq:confirm}.
In other words, each milestone confirms a part of history. 
Therefore, as long as the peers reach a consensus on all milestones, they reach a consensus on the entire history.
To achieve the objective, we have to put all milestones in a chain structure by requiring all blocks to have a pointer pointing to a previous milestone.
This is because when preparing a block, in particular when setting the pointer $\idm$, there is no way of knowing whether or not that block will be a milestone until cryptographic puzzle is solved.
If the block turns out to be a milestone, we want to make sure it connects to a previous milestone via the pointer $\idm$.

Let $\sdag$ be a collection of blocks including the genesis $\gen$, such that each non-genesis block satisfies \eqref{eq:idp}--\eqref{eq:idm}, and
\begin{equation}
  \label{dag:close}
  \forall\ \blk\in\sdag, 
  H(\blk') = \blk.\texttt{id}_{\texttt{key}},\ \texttt{key}\in\{\texttt{prev}, \texttt{ms}, \texttt{tip}\}
  \implies 
  \blk'\in \sdag.
\end{equation}
In other words, all blocks to which $\blk\in\sdag$ points are also in $\sdag$.
Essentially, $\sdag$ is a directed graph if we regard blocks as vertices and pointers as directed edges.
We say there exists a \emph{path} from $\blk$ to $\blk'$ if starting at block $\blk$ we can follow a consistently-directed sequence of edges to reach block $\blk'$.
Such a graph is called \textit{acyclic} if for any block $\blk$ in $\sdag$ there is no directed path from $\blk$ to itself.

A block is \emph{syntactically valid} if its format satisfies \eqref{eq:block} and it matches its identity. 
This is similar to the integrity check of a block in Bitcoin.
A DAG $\sdag$ is \emph{syntactically valid} if all of its blocks are syntactically valid, and it is acyclic and satisfies \eqref{dag:close}.
A block $\blk$ is \emph{syntactically valid} for the sDAG $\sdag$ if $\sdag\cup\{\blk\}$ is syntactically valid.
A peer's first task is to ensure his local DAG is syntactically valid.
Such a check protects the system from being flooded with invalid blocks.

\begin{figure}[htbp!]
  \centering
  \begin{tikzpicture}[scale=0.9]
    \node[draw,rectangle,minimum size=0.32cm] (a5) at (0.5,0){};
    \node[draw,rectangle,minimum size=0.32cm] (b5) at (4.3,0){};
    \node[draw,rectangle,minimum size=0.32cm] (c5) at (9.8,0){};
    \node[draw,rectangle, fill=red,draw=red,minimum size=0.32cm] (d5) at (14,0) {};
    \node[draw,rectangle,minimum size=0.32cm] (a4) at (1.5,1){};
    \node[draw,rectangle,minimum size=0.32cm] (b4) at (5.3,1){};
    \node[draw,rectangle,fill= red,draw=red,minimum size=0.32cm] (c4) at (10.8,1) {};
    \node[draw,rectangle,minimum size=0.32cm] (d4) at (12,1){};
    \node[draw,rectangle,minimum size=0.32cm] (a3) at (2.5,2){};
    \node[draw,rectangle, fill=red,draw=red, minimum size=0.32cm] (b3) at (6.5,2){};
    \node[draw,rectangle,minimum size=0.32cm] (c3) at (8.5,2){};
    \node[draw,rectangle,minimum size=0.32cm] (d3) at (13,2){};
    \node[draw,rectangle,fill= red,draw=red,minimum size=0.32cm] (a2) at (0,3) {};
    \node[draw,rectangle,minimum size=0.32cm] (b2) at (5.8,3){};
    \node[draw,rectangle, fill=blue,draw=blue,minimum size=0.32cm] (c2) at (10.5,3){};
    \node[draw,rectangle,minimum size=0.32cm] (d2) at (12.3,3){};
    \node[draw,rectangle,fill= red,draw=red,minimum size=0.32cm] (a1) at (3.5,4) {};
    \node[draw,rectangle,minimum size=0.32cm] (b1) at (9.2,4) {};
    \node[draw,rectangle,minimum size=0.32cm] (c1) at (14,4) {};
    \node[draw,rectangle,minimum size=0.7cm] (o) at (-2,2) {\texttt{O}};

    \node[inner sep=0,minimum size=0] at (0,0) (q) {}; 
    \node[inner sep=0,minimum size=0] at (0,1) (r) {};
    \node[inner sep=0,minimum size=0] at (0,3) (s) {};
    \node[inner sep=0,minimum size=0] at (0,4) (t) {};
    \node [font=\fontsize{7}{7}\selectfont] at (15,0) {miner 5};
    \node [font=\fontsize{7}{7}\selectfont] at (15,1) {miner 4};
    \node [font=\fontsize{7}{7}\selectfont] at (15,2) {miner 3};
    \node [font=\fontsize{7}{7}\selectfont] at (15,3) {miner 2};
    \node [font=\fontsize{7}{7}\selectfont] at (15,4) {miner 1};
    \draw [color=red,>=stealth,->,dashed] (a2)--(o);
    \draw [color=red,>=stealth,->] (b3)--(a1);
    \draw [color=red,>=stealth,->] (a1)--(a2);
    \draw [color=blue,>=stealth,->] (c2)--(b3);
    \draw [color=red,>=stealth,->] (c4)--(b3);
    \draw [color=red,>=stealth,->] (d5)--(c4);
    \draw [color=black,-](a4)-- (0,1);
    \draw [color=black,-](a1)-- (0,4);
    \draw [color=black,-](a5)-- (0,0);
    \draw [color=black,>=stealth,->](b5)-- (a5);
    \draw [color=black,>=stealth,->](c5)-- (b5);
    \draw [color=black,>=stealth,->](d5)-- (c5);
    \draw [color=black,>=stealth,->](b4)-- (a4);
    \draw [color=black,>=stealth,->](c4)-- (b4);
    \draw [color=black,>=stealth,->](d4)-- (c4);
    \draw [color=black,-](a3)-- (0,2);
    \draw [color=black,>=stealth,->](b3)-- (a3);
    \draw [color=black,>=stealth,->](c3)-- (b3);
    \draw [color=black,>=stealth,->](d3)-- (c3);
    \draw [color=black,>=stealth,->](b2)-- (a2);
    \draw [color=black,>=stealth,->](c2)-- (b2);
    \draw [color=black,>=stealth,->](d2)-- (c2);
    \draw [color=black,>=stealth,->](b1)-- (a1);
    \draw [color=black,>=stealth,->](c1)-- (b1);
    \draw [color=black,>=stealth,->,dashed](0,0)-- (o);
    \draw [color=black,>=stealth,->](a4)-- (a5);
    \draw [color=black,>=stealth,->](a4)-- (a2);
    \draw [color=black,>=stealth,->,dashed](0,2)-- (o);
    \draw [color=black,>=stealth,->,dashed](0,1)-- (o);
    \draw [color=black,>=stealth,->,dashed] (0,4)-- (o);
    \draw [color=black,>=stealth,->](a3)-- (a2);
    \draw [color=black,>=stealth,->](a3)-- (a4);
    \draw [color=black,>=stealth,->](a1)-- (a3);
    \draw [color=black,>=stealth,->](b5)-- (a2);
    \draw [color=black,>=stealth,->](b5)-- (a3);
    \draw [color=black,>=stealth,->](b4)-- (b5);
    \draw [color=black,>=stealth,->](b4)-- (a1);
    \draw [color=black,>=stealth,->](b3)-- (b2);
    \draw [color=black,>=stealth,->](b2)-- (a1);
    \draw [color=black,>=stealth,->](b2)-- (b4);
    \draw [color=black,>=stealth,->](c3)-- (b2);
    \draw [color=black,>=stealth,->](c5)-- (b3);
    \draw [color=black,>=stealth,->](c5)-- (c3);
    \draw [color=black,>=stealth,->](c4)-- (c5);
    \draw [color=black,>=stealth,->](b1)-- (c3);
    \draw [color=black,>=stealth,->](c2)-- (b2);
    \draw [color=black,>=stealth,->](b1)-- (b3);
    \draw [color=black,>=stealth,->](d4)-- (c5);
    \draw [color=black,>=stealth,->](c2)-- (b1);
    \draw [color=black,>=stealth,->](d2)-- (b1);
    \draw [color=black,>=stealth,->](c1)-- (d2);
    \draw [color=black,>=stealth,->](c1)-- (c2);
    \draw [color=black,>=stealth,->](d5)-- (d3);
    \draw [color=black,>=stealth,->](d3)-- (c4);
    \draw [color=black,>=stealth,->](d3)-- (d4);
    \end{tikzpicture}
  \caption{DAG structure for 5 miners with milestone forks.}
  \label{fig:dag}
\end{figure}
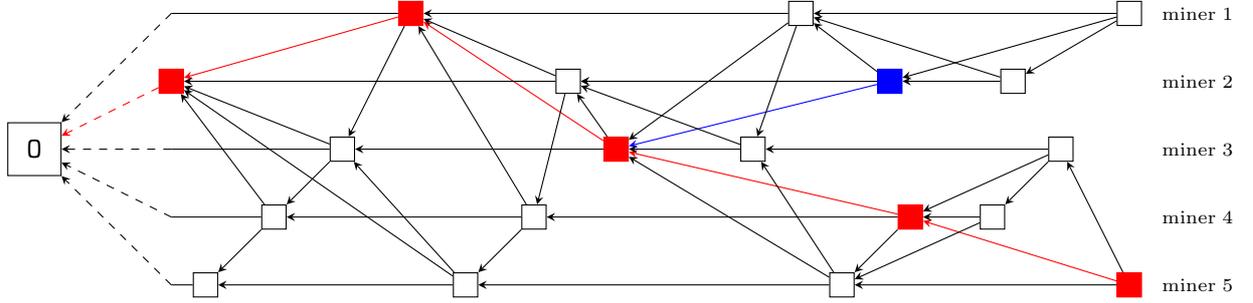

It will become self-evident that under our protocol, the milestone tree is essentially the tree occurring in a blockchain with forks. 
The milestone tree plays an essential role in connecting all of the peer chains, while the pointers $\idt$ further enhance the connectivity of the DAG
Figure~\ref{fig:dag} provides an illustration of such a structure.

We now define the \textit{height} of milestone blocks in a structured DAG. 
The height of the genesis $\height(\gen)=0$, and for a milestone block $\blk$, i.e., $H(\blk)<p\trgt$, the height is defined as
\begin{equation}
  \label{eq:height_ms}
  \height(\blk) = \height(\blk_m) +1,
  \quad 
  \text{where } H(\blk)<p\trgt
  \text{ and } \blk.\idm = H(\blk_m).
\end{equation}
The genesis and all milestone blocks in $\sdag$ form a tree where the depth of each milestone block is its height.
The \emph{leaf set} of all milestone blocks is 
\begin{equation}
  \begin{split}
      \leaf_m(\sdag) = \{\blk_m\in\sdag:\ 
      &H(\blk_m)<p\trgt \text{ and }\\ 
      &\nexists\ \blk'_m\in\sdag \text{ s.t.\ }H(\blk'_m)<p\trgt \text{ and } \blk'_m.\idm=H(\blk_m)\}.
  \end{split}
\end{equation}
For a milestone block $\blk_m$ in the leaf set with $\height(\blk_m) = n$, following the pointers $\idm$ we can find a sequence of blocks
\begin{equation}
  \blk_{m,n}=\blk_m, \blk_{m,n-1}, \blk_{m,n-2}, ...., \blk_{m,1}, \blk_{m,0}=\gen, 
\end{equation}
such that $H(\blk_{m,k})<p\trgt$ and $\blk_{m,k}.\idm = H(\blk_{m,k-1})$ for all $k=1,2,\ldots,n$.
We call such a sequence the \textit{milestone chain }for block $\blk_m$.  
So each block in the leaf set of all milestone blocks represents a milestone chain, whose length is equal to the height of that block. 
We define the height of the DAG to be the length of the longest chain(s),
\begin{equation}
  \height(\sdag) = \max\{
    \height(\blk): 
    \blk \in \leaf_m(\sdag)
  \}.
\end{equation}
Note that multiple longest chains might exist. 
To choose a longest chain, we just need to choose a block with the largest height in the leaf set. 
The $n$th milestone is defined to be the block on the longest milestone chain with height $n$.

\begin{figure}[htbp!]
  \centering
  \begin{tikzpicture}[scale=0.9]
    \shade[top color=gray, bottom color = gray, opacity=1]
    (-1,-0.2) rectangle (0.2,4.2);
    \shade[top color=gray!70, bottom color = gray!70, opacity=1]
    (0.3,-0.2) rectangle (3.9,4.2);
    \shade[top color=gray!50, bottom color = gray!50, opacity=1]
    (4.1,-0.2) rectangle (7,4.2);
    \shade[top color=gray!30, bottom color = gray!30, opacity=1]
    (7.6,-0.2) rectangle (11.1, 2.2);
    \shade[top color=gray!15, bottom color = gray!15, opacity=1]
    (11.6,-0.2) rectangle (14.3, 2.2);
    \shade[top color=orange!40, bottom color = orange!40, opacity=1]
    (8.6,2.8) rectangle (14.3,4.2);
    \node[draw,rectangle,minimum size=0.32cm] (a5) at (0.5,0){};
    \node[draw,rectangle,minimum size=0.32cm] (b5) at (4.3,0){};
    \node[draw,rectangle,minimum size=0.32cm] (c5) at (9.8,0){};
    \node[draw,rectangle, fill =red, draw=red,minimum size=0.32cm] (d5) at (14,0) {};
    \node[draw,rectangle,minimum size=0.32cm] (a4) at (1.5,1){};
    \node[draw,rectangle,minimum size=0.32cm] (b4) at (5.3,1){};
    \node[draw,rectangle,fill= red, draw=red,minimum size=0.32cm] (c4) at (10.8,1) {};
    \node[draw,rectangle,minimum size=0.32cm] (d4) at (12,1){};
    \node[draw,rectangle,minimum size=0.32cm] (a3) at (2.5,2){};
    \node[draw,rectangle, fill=red,draw=red, minimum size=0.32cm] (b3) at (6.5,2){};
    \node[draw,rectangle,minimum size=0.32cm] (c3) at (8.5,2){};
    \node[draw,rectangle,minimum size=0.32cm] (d3) at (13,2){};
    \node[draw,rectangle,fill= red,draw=red,minimum size=0.32cm] (a2) at (0,3) {};
    \node[draw,rectangle,minimum size=0.32cm] (b2) at (5.8,3){};
    \node[draw,rectangle, fill=blue,draw=blue,minimum size=0.32cm] (c2) at (10.5,3){};
    \node[draw,rectangle,minimum size=0.32cm] (d2) at (12.3,3){};
    \node[draw,rectangle,fill= red,draw=red, minimum size=0.32cm] (a1) at (3.5,4) {};
    \node[draw,rectangle,minimum size=0.32cm] (b1) at (9.2,4) {};
    \node[draw,rectangle,minimum size=0.32cm] (c1) at (14,4) {};
    \node[draw,rectangle,minimum size=0.7cm] (o) at (-2,2) {\texttt{O}};

    \node[inner sep=0,minimum size=0] at (0,0) (q) {}; 
    \node[inner sep=0,minimum size=0] at (0,1) (r) {};
    \node[inner sep=0,minimum size=0] at (0,3) (s) {};
    \node[inner sep=0,minimum size=0] at (0,4) (t) {};
    \node [font=\fontsize{7}{7}\selectfont] at (15,0) {miner 5};
    \node [font=\fontsize{7}{7}\selectfont] at (15,1) {miner 4};
    \node [font=\fontsize{7}{7}\selectfont] at (15,2) {miner 3};
    \node [font=\fontsize{7}{7}\selectfont] at (15,3) {miner 2};
    \node [font=\fontsize{7}{7}\selectfont] at (15,4) {miner 1};
    \draw [color=red,>=stealth,->,dashed] (a2)--(o);
    \draw [color=red,>=stealth,->] (b3)--(a1);
    \draw [color=red,>=stealth,->] (a1)--(a2);
    \draw [color=blue,>=stealth,->] (c2)--(b3);
    \draw [color=red,>=stealth,->] (c4)--(b3);
    \draw [color=red,>=stealth,->] (d5)--(c4);
    \draw [color=black,-](a4)-- (0,1);
    \draw [color=black,-](a1)-- (0,4);
    \draw [color=black,-](a5)-- (0,0);
    \draw [color=black,>=stealth,->](b5)-- (a5);
    \draw [color=black,>=stealth,->](c5)-- (b5);
    \draw [color=black,>=stealth,->](d5)-- (c5);
    \draw [color=black,>=stealth,->](b4)-- (a4);
    \draw [color=black,>=stealth,->](c4)-- (b4);
    \draw [color=black,>=stealth,->](d4)-- (c4);
    \draw [color=black,-](a3)-- (0,2);
    \draw [color=black,>=stealth,->](b3)-- (a3);
    \draw [color=black,>=stealth,->](c3)-- (b3);
    \draw [color=black,>=stealth,->](d3)-- (c3);
    \draw [color=black,>=stealth,->](b2)-- (a2);
    \draw [color=black,>=stealth,->](c2)-- (b2);
    \draw [color=black,>=stealth,->](d2)-- (c2);
    \draw [color=black,>=stealth,->](b1)-- (a1);
    \draw [color=black,>=stealth,->](c1)-- (b1);
    \draw [color=black,>=stealth,->,dashed](0,0)-- (o);
    \draw [color=black,>=stealth,->](a4)-- (a5);
    \draw [color=black,>=stealth,->](a4)-- (a2);
    \draw [color=black,>=stealth,->,dashed](0,2)-- (o);
    \draw [color=black,>=stealth,->,dashed](0,1)-- (o);
    \draw [color=black,>=stealth,->,dashed] (0,4)-- (o);
    \draw [color=black,>=stealth,->](a3)-- (a2);
    \draw [color=black,>=stealth,->](a3)-- (a4);
    \draw [color=black,>=stealth,->](a1)-- (a3);
    \draw [color=black,>=stealth,->](b5)-- (a2);
    \draw [color=black,>=stealth,->](b5)-- (a3);
    \draw [color=black,>=stealth,->](b4)-- (b5);
    \draw [color=black,>=stealth,->](b4)-- (a1);
    \draw [color=black,>=stealth,->](b3)-- (b2);
    \draw [color=black,>=stealth,->](b2)-- (a1);
    \draw [color=black,>=stealth,->](b2)-- (b4);
    \draw [color=black,>=stealth,->](c3)-- (b2);
    \draw [color=black,>=stealth,->](c5)-- (b3);
    \draw [color=black,>=stealth,->](c5)-- (c3);
    \draw [color=black,>=stealth,->](c4)-- (c5);
    \draw [color=black,>=stealth,->](b1)-- (c3);
    \draw [color=black,>=stealth,->](c2)-- (b2);
    \draw [color=black,>=stealth,->](b1)-- (b3);
    \draw [color=black,>=stealth,->](d4)-- (c5);
    \draw [color=black,>=stealth,->](c2)-- (b1);
    \draw [color=black,>=stealth,->](d2)-- (b1);
    \draw [color=black,>=stealth,->](c1)-- (d2);
    \draw [color=black,>=stealth,->](c1)-- (c2);
    \draw [color=black,>=stealth,->](d5)-- (d3);
    \draw [color=black,>=stealth,->](d3)-- (c4);
    \draw [color=black,>=stealth,->](d3)-- (d4);
    \node [font=\fontsize{8}{8}\selectfont] at (-0.4,-0.5) {$k-1$};
    \node [font=\fontsize{8}{8}\selectfont] at (2,-0.5) {$k$};
    \node [font=\fontsize{8}{8}\selectfont] at (5.5,-0.5) {$k+1$};
    \node [font=\fontsize{8}{8}\selectfont] at (9.3,-0.5) {$k+2$};
    \node [font=\fontsize{8}{8}\selectfont] at (13,-0.5) {$k+3$};
    \node [font=\fontsize{8}{8}\selectfont] at (11.5,4.4) {Pending Set};
  \end{tikzpicture}
  \caption{An Illustration of Level Sets and the Pending Set}
  \label{fig:dag-levelset}
\end{figure}
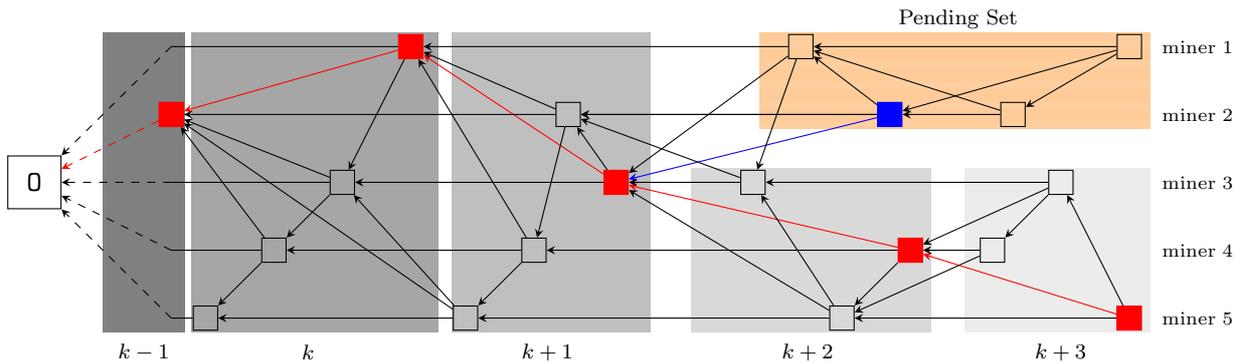

For any milestone block $\blk_m\in\sdag$, we define the \emph{DAG confirmed} by the milestone $\blk_m$ to be
\begin{equation}
  \label{eq:confirm}
  \confirm(\blk_m) = \{
    \blk\in\sdag: 
    \text{ there exits a path from }\blk \text{ to } \blk_m
    \}
    \cup\{\blk_m\}. 
\end{equation}
If $\blk'_m\in\sdag$ is the milestone or genesis that immediately preceding $\blk_m$, i.e.\ $\blk_m.\idm = H(\blk'_m)$, then we define the \emph{level set} to be
\begin{equation}
  \label{eq:level-set-ms}
  \lev(\blk_m,\blk'_m) = \confirm(\blk_m)\setminus\confirm(\blk'_m). 
\end{equation}
With slight extension of notation, we define $\confirm(k)=\confirm(\blk_{m,k})$ to be the DAG confirmed by the $k$th milestone and $\lev(k)=\confirm(k)\setminus\confirm(k-1)$ to be the $k$th level set, with $\confirm(0)=\{\gen\}$. 
In Figure~\ref{fig:dag-levelset}, level sets $\lev(k)$, $\lev(k+1)$, $\lev(k+2)$ and $\lev(k+3)$ are shaded in gray.
The \emph{pending set} is defined to be the set of all blocks that have not yet been confirmed by any milestone on the longest milestone chain.
It is worth pointing out that the blue block is also a milestone block. This milestone forked from the $(k+1)$th milestone due to either network synchronization delay or attack. The blue and red blocks in Figure~\ref{fig:dag-levelset} form the milestone tree above defined. The blue block and the current $(k+2)$th milestone will compete for confirmations by future milestones based on Nakamoto consensus. If the blue blocks fails, i.e.\ all honest peers thinks it is a fork, it will be treated as a regular block in Section~\ref{sec:DAG-to-Ledger} when we building the ledger.

\section{Protocol for Maintaining a Local DAG}
\label{sec:protocol}

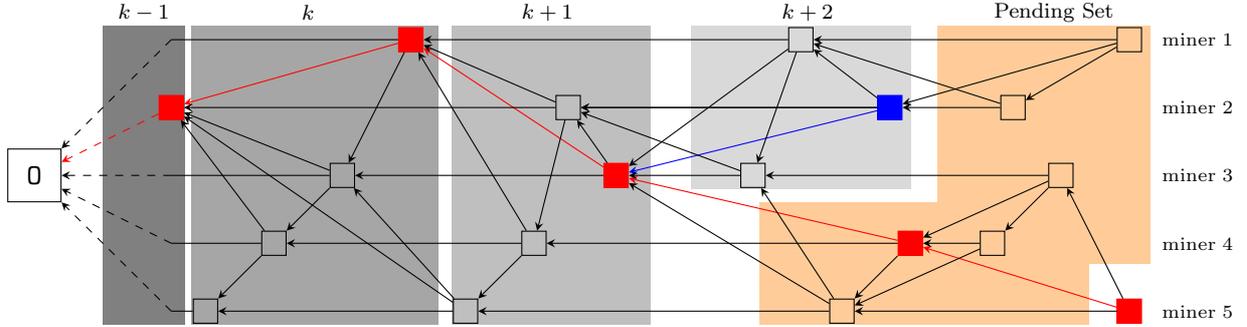
\begin{figure}[htbp!]
  \centering
  \begin{tikzpicture}[scale=0.9]
    \shade[top color=gray, bottom color = gray, opacity=1]
  (-1,-0.2) rectangle (0.2,4.2);
   \shade[top color=gray!70, bottom color = gray!70, opacity=1]
  (0.3,-0.2) rectangle (3.9,4.2);
   \shade[top color=gray!50, bottom color = gray!50, opacity=1]
  (4.1,-0.2) rectangle (7,4.2);
   \shade[top color=gray!30, bottom color = gray!30, opacity=1]
  (7.6,4.2) rectangle (10.8, 1.8);
 \shade[top color=orange!40, bottom color = orange!40, opacity=1]
  (8.6,-0.2) rectangle (13.4,1.6);
  \shade[top color=orange!40, bottom color = orange!40, opacity=1]
  (11.2,0.7) rectangle (14.3,4.2);     
    \node[draw,rectangle,minimum size=0.32cm] (a5) at (0.5,0){};
    \node[draw,rectangle,minimum size=0.32cm] (b5) at (4.3,0){};
    \node[draw,rectangle,minimum size=0.32cm] (c5) at (9.8,0){};
    \node[draw,rectangle, fill =red,draw=red, minimum size=0.32cm] (d5) at (14,0) {};
    \node[draw,rectangle,minimum size=0.32cm] (a4) at (1.5,1){};
    \node[draw,rectangle,minimum size=0.32cm] (b4) at (5.3,1){};
    \node[draw,rectangle,fill= red,draw=red, minimum size=0.32cm] (c4) at (10.8,1) {};
    \node[draw,rectangle,minimum size=0.32cm] (d4) at (12,1){};
    \node[draw,rectangle,minimum size=0.32cm] (a3) at (2.5,2){};
    \node[draw,rectangle, fill=red,draw=red,  minimum size=0.32cm] (b3) at (6.5,2){};
    \node[draw,rectangle,minimum size=0.32cm] (c3) at (8.5,2){};
    \node[draw,rectangle,minimum size=0.32cm] (d3) at (13,2){};
    \node[draw,rectangle,fill= red,draw=red, minimum size=0.32cm] (a2) at (0,3) {};
    \node[draw,rectangle,minimum size=0.32cm] (b2) at (5.8,3){};
    \node[draw,rectangle, fill=blue,draw=blue, minimum size=0.32cm] (c2) at (10.5,3){};
    \node[draw,rectangle,minimum size=0.32cm] (d2) at (12.3,3){};
    \node[draw,rectangle,fill= red,draw=red,minimum size=0.32cm] (a1) at (3.5,4) {};
    \node[draw,rectangle,minimum size=0.32cm] (b1) at (9.2,4) {};
    \node[draw,rectangle,minimum size=0.32cm] (c1) at (14,4) {};
    \node[draw,rectangle,minimum size=0.7cm] (o) at (-2,2) {\texttt{O}};

    \node[inner sep=0,minimum size=0] at (0,0) (q) {}; 
    \node[inner sep=0,minimum size=0] at (0,1) (r) {};
    \node[inner sep=0,minimum size=0] at (0,3) (s) {};
    \node[inner sep=0,minimum size=0] at (0,4) (t) {};
    \node [font=\fontsize{7}{7}\selectfont] at (15,0) {miner 5};
    \node [font=\fontsize{7}{7}\selectfont] at (15,1) {miner 4};
    \node [font=\fontsize{7}{7}\selectfont] at (15,2) {miner 3};
    \node [font=\fontsize{7}{7}\selectfont] at (15,3) {miner 2};
    \node [font=\fontsize{7}{7}\selectfont] at (15,4) {miner 1};
    \draw [color=red,>=stealth,->,dashed] (a2)--(o);
    \draw [color=red,>=stealth,->] (b3)--(a1);
    \draw [color=red,>=stealth,->] (a1)--(a2);
    \draw [color=blue,>=stealth,->] (c2)--(b3);
    \draw [color=red,>=stealth,->] (c4)--(b3);
    \draw [color=red,>=stealth,->] (d5)--(c4);
    \draw [color=black,-](a4)-- (0,1);
    \draw [color=black,-](a1)-- (0,4);
    \draw [color=black,-](a5)-- (0,0);
    \draw [color=black,>=stealth,->](b5)-- (a5);
    \draw [color=black,>=stealth,->](c5)-- (b5);
    \draw [color=black,>=stealth,->](d5)-- (c5);
    \draw [color=black,>=stealth,->](b4)-- (a4);
    \draw [color=black,>=stealth,->](c4)-- (b4);
    \draw [color=black,>=stealth,->](d4)-- (c4);
    \draw [color=black,-](a3)-- (0,2);
    \draw [color=black,>=stealth,->](b3)-- (a3);
    \draw [color=black,>=stealth,->](c3)-- (b3);
    \draw [color=black,>=stealth,->](d3)-- (c3);
    \draw [color=black,>=stealth,->](b2)-- (a2);
    \draw [color=black,>=stealth,->](c2)-- (b2);
    \draw [color=black,>=stealth,->](d2)-- (c2);
    \draw [color=black,>=stealth,->](b1)-- (a1);
    \draw [color=black,>=stealth,->](c1)-- (b1);
    \draw [color=black,>=stealth,->,dashed](0,0)-- (o);
    \draw [color=black,>=stealth,->](a4)-- (a5);
    \draw [color=black,>=stealth,->](a4)-- (a2);
    \draw [color=black,>=stealth,->,dashed](0,2)-- (o);
    \draw [color=black,>=stealth,->,dashed](0,1)-- (o);
    \draw [color=black,>=stealth,->,dashed] (0,4)-- (o);
    \draw [color=black,>=stealth,->](a3)-- (a2);
    \draw [color=black,>=stealth,->](a3)-- (a4);
    \draw [color=black,>=stealth,->](a1)-- (a3);
    \draw [color=black,>=stealth,->](b5)-- (a2);
    \draw [color=black,>=stealth,->](b5)-- (a3);
    \draw [color=black,>=stealth,->](b4)-- (b5);
    \draw [color=black,>=stealth,->](b4)-- (a1);
    \draw [color=black,>=stealth,->](b3)-- (b2);
    \draw [color=black,>=stealth,->](b2)-- (a1);
    \draw [color=black,>=stealth,->](b2)-- (b4);
    \draw [color=black,>=stealth,->](c3)-- (b2);
    \draw [color=black,>=stealth,->](c5)-- (b3);
    \draw [color=black,>=stealth,->](c5)-- (c3);
    \draw [color=black,>=stealth,->](c4)-- (c5);
    \draw [color=black,>=stealth,->](b1)-- (c3);
    \draw [color=black,>=stealth,->](b1)-- (b3);
    \draw [color=black,>=stealth,->](c2)-- (b2);
    \draw [color=black,>=stealth,->](d4)-- (c5);
    \draw [color=black,>=stealth,->](c2)-- (b1);
    \draw [color=black,>=stealth,->](d2)-- (b1);
    \draw [color=black,>=stealth,->](c1)-- (d2);
    \draw [color=black,>=stealth,->](c1)-- (c2);
    \draw [color=black,>=stealth,->](d5)-- (d3);
    \draw [color=black,>=stealth,->](d3)-- (c4);
    \draw [color=black,>=stealth,->](d3)-- (d4);
    \node [font=\fontsize{8}{8}\selectfont] at (-0.4,4.4) {$k-1$};
    \node [font=\fontsize{8}{8}\selectfont] at (2,4.4) {$k$};
    \node [font=\fontsize{8}{8}\selectfont] at (5.5,4.4) {$k+1$};
    \node [font=\fontsize{8}{8}\selectfont] at (9.3,4.4) {$k+2$};
    \node [font=\fontsize{8}{8}\selectfont] at (12.9,4.4) {Pending Set};
  \end{tikzpicture}
  \caption{Level Sets from a Different Perspective}
\label{fig:dag-levelset-fork}
\end{figure}

The structured DAG is collectively created and maintained in a distributed fashion by all peers over a peer-to-peer network following a protocol.
Peers following this protocol are said to be \emph{honest} and peers not following this protocol are said to be \emph{malicious}.
In a decentralized system, each peer will have his own local DAG, which can be different from the local DAG of another peer due to issues like network synchronization delay and malicious attacks. This can be illustrated by comparing Figure~\ref{fig:dag-levelset} and Figure~\ref{fig:dag-levelset-fork}. 
Suppose the due to network synchronization, an honest peer has not received the latest milestone created by miner~5. He may then think that the blue milestone is the $(k+2)$th on the longest chain. In this case, his view is depicted in Figure~\ref{fig:dag-levelset-fork}.

For a peer $a$, denote his local DAG by $\sdag_a$, which is evolving he creates and receives blocks. 
All miners start their peer chains from the genesis block $\gen$. 
We now describe the protocol to be followed by each peer/miner.
The objective for all honest peers to agree upon the same DAG (see Section~\ref{sec:concensus}), and consequently to use the same set of data to construct the public ledger.
We will present our protocol in the setting of a cryptocurrency application for concreteness and clarity, although the protocol can potentially be generalized to other applications. 
To this end, we first discuss the mempool, which is the buffer holding all pending transactions to be processed.

\subsection{Mempool Transaction Assignment}
\label{sec:mempool}

When someone wants to initiate a transaction, he will broadcast it to a few connected peers. This is implemented via an inventory message and getdata approach in Bitcoin for example.
If a peer considers the transaction valid after receiving it, he will also broadcast the transaction to all of his peers. 
A peer also need to remove a transaction from his local mempool once he find out that transaction has already been put in a block, created either by himself or other peers. 
In such a way, all peers will locally have a buffer of all outstanding transactions waiting to be processed. 

The purpose of expanding the chain structure to a DAG structure is to allow parallelism for scaling up the capacity.
However, with the current mempool design, it is highly possible that the same transaction, especially when it is associated with a high transaction fee, will be processed by multiple miners due to network broadcast delay. 
Consequently, the same transaction may end up in multiple blocks. 
Although only one block will be valid according to our local algorithm of constructing the ledger from the local DAG, much capacity would go to waste. 
If someone with malicious intent wants to ruin the network by wasting a large portion of the capacity, he could broadcast transactions with attractive fees in rapid succession. 
Such an attack would be much less costly than owning and using a certain percentage of the hashing power on the one hand, and can effectively reduce the capacity on the other hand.

Therefore, we need to design a mechanism to effectively reduce the chance of collision, i.e., the chance of one transaction being processed by multiple miners.
The basic idea is to dynamically limit the number of transactions a miner can process at any time by using the hash result $.H(H(\blk_i),\tx)$ as a distance between transaction \tx{} and the miner's state $\blk_i$, i.e. the most recent block on miner~$i$'s peer chain.
The peer chain can also give a clear count of the proportion of blocks created by this miner in any level set, and thus it provides a reference of the miner's hashing power. 
Suppose miner~$i$ possesses $q_i\in[0,1]$ proportion of the total hashing power.
We say that transaction \tx{} is workable for miner~$i$ only if the distance
\begin{equation}
  \label{eq:tx_distance}
  .H(H(\blk_i),\tx)\leq cq_i, 
\end{equation}
where $c$ is parameter to be specified based on some careful analysis in Section~\ref{sec:analysis}.
This means that a transaction has a probability of only $\min(cq_i, 1)$ to be workable for miner~$i$ at any time.

We now provide some simple calculation to give some idea.
First, miner~$i$ can surely process all transactions if $c$ is chosen such that $cq_i\ge1$. So we just limit our choice of $c$ to the range $(0,1/\max_iq_i)$. 
Under this assumption, we have 
\begin{equation}
  \label{eq:noworkprob}
  \begin{split}
  \prob(\tx \ \text{is not workable for any miner}) &= \prod_{i=1}^n(1-cq_i) \\
  &\leq (\frac{\sum_{i=1}^n(1-cq_i)}{n})^n \\
  &= \Big(1-\frac{c}{n}\Big)^n \\
  &\approx e^{-c},
  \end{split}
\end{equation}
as $n$ becomes large.
So intuitively, making $c$ small cause a transaction hard to be processed by miners in the network.
Here we want to estimate the probability that a transaction can be processed by two or more miners. To do so, we need to compute
\begin{equation}
  \label{eqn:workable = 1}
  \begin{split}
  \prob(\tx \text{ is workable for exactly one miner}) 
  &= \sum_{i=1}^{n} cq_i\prod_{j \ne i}(1-cq_j)\\
  &\approx e^{-c} \sum_{i=1}^{n}\frac{cq_i}{1-cq_i}\\
  &\approx ce^{-c},
  \end{split}
\end{equation}
as $n$ becomes large.
The last estimate in the above is under the assumption that no miner owns a significant portion hashing power thus $1-cq_i$ can be approximated by 1 when $n$ is large.
In this way, we obtain a neat answer and we can see clearly how the parameter $c$ plays a role.
Summarizing the above two estimates, the probability of collision is given by
\begin{equation}
  \label{eq:collision}
  \prob(\tx \text{ is workable by more than one miners}) 
  \approx 1- e^{-c} - ce^{-c}.
\end{equation}
Making $c$ smaller helps to reduce collision. 
To give some quantitative idea, simply setting $c=0.1$ will make the collision probability less than $0.5\%$.

Note that the set of workable transactions for each miner is dynamic as the miner state changes every time the miner creates a new block.
So how long a transaction has to wait in the mempool and how much capacity will be wasted due to collision require detailed analysis via a modeling approach in Sections~\ref{sec:throughput} and \ref{sec:latency}.
Through such an analysis, we can choose the parameter $c$ wisely to strike a balance between the wasted capacity and waiting time.

\subsection{Receiving a Block}
A peer may be new to the network or has been offline for a while. In this case, the peer should first learn from his connected peers the height of the longest milestone chain. If the height of his local DAG is much smaller, he should start downloading the missing level sets.
Once a peer's local height is close to the heights of his connected peers, e.g., within a threshold specified in our code implementation, the peer can start to receive broadcasted newly mined blocks.
Suppose peer~$a$ receives a block $\blk$ through broadcast. 
It is possible that he does not have some blocks to which $\blk$ either directly or indirectly points due to synchronization delay.
He should start a process to \emph{solidify} the block $\blk$ by asking his peers for the missing blocks from its peers.
Once he has obtained all the missing blocks, peer~$a$ needs perform topological sorting (e.g.\ Section~2.2.3 in \cite{KnuthTAOCP1}) of these blocks, so that blocks can be added sequentially to the peer's local DAG.

We now describe how to add a block $\blk$, where the blocks to which it points are already in the DAG $\sdag_a$.
First, check if block $\blk$ is syntactically valid for $\sdag_a=\confirm(\blk_m)$. If not, discarded the it immediately; otherwise update the local DAG $\sdag_a$ by letting 
\begin{equation*}
  \sdag_a := \sdag_a \cup \{\blk\}  
\end{equation*}
and relay the block $\blk$ to connected peers.
If block $\blk$ contains a transaction that is in the peer's mempool, the peer removes the transaction from his mempool.
Note that we do not require the peer to verify the transaction in block $\blk$ at this stage. 
Taking Bitcoin for example, the verification requires verifying signatures and checking that all inputs of the transaction correspond to some unspent outputs in the public ledger according to the current local DAG, and ensuring no double spending occurs among all transactions.
We postpone such verification to the time when we convert the local DAG to a ledger, primarily because it takes time, in particular the secure latency (Section~\ref{sec:latency}), to reach consensus on the DAG.

The primary concern is whether block $\blk$ is a forked milestone or not.
In other words, we need to consider what kind of block it is. 
If block $\blk$ is regular, do nothing. 
If block $\blk$ is a milestone, we need to compare the height of $\sdag_a$ with $\height(\blk)$. If $\height(\blk)>\height(\blk_m)$, then set
\begin{equation*}
  \sdag_a:=\confirm(\blk).  
\end{equation*}
This means a major update of the $\sdag_a$ since the peer needs to switch from the current milestone chain to the newly discovered longest milestone chain in the local DAG.

\subsection{Creating a block}
If a miner wants to create a block $\blk$, and have it accepted by all other peers as part of their local DAGs, the actions to be taken are as follows.
The first thing the miner needs to do is to find a transaction that he can process from his mempool as required in by \eqref{eq:tx_distance}, and put this transaction in a block.
We also require the selected transaction to be compatible with the miner's ledger, i.e.\ the input(s) for this transaction should be in the unspent outputs of the ledger constructed based on the DAG $\confirm(\blk_m)$, where $\blk_m$ is the highest milestone in its local DAG.

Next, the miner must prepare three pointers so that the block will be syntactically valid.
The steps are as follows
\begin{enumerate}
  \item Pick the highest milestone $\blk_m$ in his local DAG $\sdag_a$ and set $\blk.\idm = H(\blk_m)$.

  Note that $\blk_m$ is the most recent milestone in the local DAG $\sdag_a$. However, due to network synchronization, $\blk_m$ is may not necessarily be the most recent milestone block in reality.

  \item Pick the most recent block $\blk'$ created by himself in $\sdag_a$ and set $\blk.\idp=H(\blk')$. If no such $\blk'$ exists, set $\blk.\idp = H(\gen)$.

  Note that block $\blk'$ and $\blk$ being consecutive blocks created by the same miner is a stronger requirement than \eqref{eq:idp}. See further discussion in the next section. 
  
  \item Define the \emph{tip set} to be the set of regular blocks to which none of the blocks in $\sdag_a$ points and are not created by the miner~$a$.
  Randomly pick a block $\blk_t$ from the tip set and set $\blk.\idt = H(\blk_t)$.
  If the tip set is $\{\gen\}$ or empty, then set $\blk.\idt = H(\gen)$.
  
  This means that the miner needs to find a regular block $\blk_t$ created by some other miner. 
  
  \item Keep applying the hash function to this block by changing the nonce untill $.H(B)<d$.

  \item Finally, the miner needs to broadcast $\blk$ through a peer-to-peer network. 
This certainly takes time, as we will discuss in the subsection on the broadcast delay assumption in Section~\ref{sec:analysis}.

\end{enumerate}

\section{Building a Public Ledger}
\label{sec:DAG-to-Ledger}

We present this section in the context of the unspent transaction output (UTXO) model of Bitcoin for concreteness.
The idea of constructing a public ledger from the DAG structure can be extended to other models.  
In the UTXO model, a \emph{transaction} essentially specifies previous transaction outputs as new transaction inputs and allocates all input values to new outputs.
Obviously, signatures are required in order to use an output as a transaction input. 
Readers can refer to Bitcoin documents for details. 
A \emph{ledger} is an ordered list of transactions, from which we can construct the set of UTXOs.
We say that a transaction is \emph{valid} if it passes the signature validation and its inputs are indeed from the UTXOs according to all preceding transactions.
A ledger is valid if all of its transactions are valid. 

Since we do not impose validity checks when receiving transactions packed in blocks, there is no guarantee that all of the transactions in a local DAG are valid. For example, it is likely two or more double-spending transactions coexist in a local DAG.
The level sets in our DAG along the longest milestone chain provide the natural ordering of level sets from low to high. 
Within each level set, we utilize a depth-first search algorithm to order all its transactions. 
In this way, as long as peers agree on the same DAG, they agree on the same way to order the same set of transactions.
We then specify an algorithm to chose a subset of this transaction following the same ordering to form the public ledger. 

Recall the definition of syntactical validity of our DAG, which every honest peer shall check when updating it. 
Although it specifies a structure in the DAG, it does not enforce hard constraints on the following requirements:
\begin{itemize}\setlength\itemsep{0pt}
  \item Each miner should keep the blocks created by himself in a chain structure without any forks.
  \item A new block shall point to the most recent milestone being created.
  \item A new block shall point to a recent regular block on another peer chain when created.
\end{itemize}
In fact, we could not enforce hard constraints in most cases since there is synchronization takes time.
For example, ideally, we would like to force each miner to connect to a block on another peer chain in the pending set. But as illustrated in Figures~\ref{fig:dag-levelset} and \ref{fig:dag-levelset-fork}, there may be no consensus on the pending set due to temporary broadcast delay. 
Also, forbidding forking of a peer chain add too many complicated rules while performing syntactical validity check and is not easy to reach consensus when there is a forking attack.
On the other side, not following the above three principles decreases the performance of our system in terms of latency and wasted capacity.
For example, connecting to a too old regular block on another peer chain contributes very little to the connectivity, thus slow down the confirmation by a milestone block.
Also, not connecting to the latest milestone creates the same issue of forking as in Bitcoin. Though eventually honest peer will reach consensus on the longest milestone chain, forking is not desirable as it is a waste of mined milestone blocks and cause delay in secure confirmation. 

\paragraph{Reward Scheme.}
We design the following reward scheme to incentivize miners to follow the above mentioned principles.
A miner should be rewarded by a fixed fee (even if the block does not contain a transaction) plus the transaction fee for creating a block.
Exactly how much a miner can get out of a block depends on three factors: the type of the block, the ultimate status of this block, and the validity of the transaction in this block.
The first two factors depends on consensus dictated by the longest milestone chain, and the last one depends on the ordering of transactions as we will explain later.

We have been using milestone and regular as block types so far. As discussed in Section~\ref{sec:dag}, all the milestone blocks form a tree. When computing the reward, we would like to treat milestone block not on the longest chain as regular blocks. 
For ease of reference, let's call regular and forked milestone blocks \emph{regular+}.
Let $r_{m}$ and $r_{n}$ denote the fixed reward for milestones on the longest milestone chain and regular+ blocks, respectively.
It is natural that $r_{m}>r_{n}$ milestone blocks are more difficult to create. As mentioned before, miners do not know the type of a block \textit{a priori} when working on it.
Let $\blk_m$ and $\blk'_m$ be two consecutive milestones on the longest milestone chain with $\blk_m.\idm = H(\blk'_m)$. 
Define
\begin{equation*}
  R(\blk_m)=r_n(|\lev(\blk_m,\blk'_m)|-1)
\end{equation*}
to be the total amount of regular+ block rewards in the level set $\lev(\blk_m,\blk'_m)$.
We summarize our reward scheme in the following table:
\begin{table}[H]
  \centering
  \begin{tabular}{|c|c|c|c|}
    \hline
    Block Type & Block Status & Transaction & Reward \\ \hline
    regular+ & on peer chain & Valid & $r_n+\tx\ \text{fee}$ \\ \hline
    regular+ & on peer chain & Invalid & $r_n$ \\ \hline
    regular+ & forked from peer chain & Valid & 0 \\ \hline
    regular+ & forked from peer chain & Invalid & 0 \\ \hline
    milestone & on the longest milestone chain & Valid & $r_m+\tx\ \text{fee} + \delta R(\blk_m)$ \\ \hline
    milestone & on the longest milestone chain & Invalid & $r_m+ \delta R(\blk_m)$ \\ \hline
  \end{tabular}
  \label{tbl:reward}
  \caption{Reward Scheme}
\end{table}

On top of fixed amount $r_m$ and the transaction fee if valid, a milestone block also brings additional reward of $\delta R(\blk_m)$. 
We can think of the parameter $\delta$ as for example $2\%$. This means that any miner who created a milestone, he will get in addition 2\% of the rewards for all the blocks that milestone confirmed in its level set.
For example, if a system is running at 1000 TPS, and milestones are generated every 10 seconds, there will be on average 10000 blocks in a level set. 
So there is quite a big bonus compared with the situation where the block, despite qualified to be a milestone, end up being a regular+ block because it is not on the longest chain. 
By appropriately chose the parameters $r_n,r_m$ and $\delta$, we hope to provide miners enough incentive to try their best to point to the most recent milestone when creating new blocks.
Since the extra bonus depend on the number of blocks in the level set, this will also incentivize a miner to point a new block to a recent regular block on another peer chain to maximize the number of blocks he can confirm. 

The economic model is out of the scope of this paper which focus on consensus. 
Under our design, one can choose to adjust $r_n$ and $r_m$ every once in a while so that the total number of issued coins is fixed like in Bitcoin. 
Alternatively, one can keep them fixed so that the total amount of currency in the system will inflate. 
There are many ways to design the financial system, which we will not discuss here. 
No matter how to choose these parameters, such a scheme together with smaller blocks which can be miner more easily will make miner~$i$'s income rate commensurate (without too much volatility) to his $q_i$ proportion of total hashing power times the rate at which new currencies are issued plus transaction fees.

Note that in the above table, forked blocks from a peer chain brings zero reward thus incentivize miners to follow the principle of maintaining the structure in peer chains.
How exactly does this work will be discussed in the following part.  

\paragraph{Registration and Redemption.}
As we can see in the designed scheme, the reward for a block cannot be recorded as a static numerical number upon the time creating the block. Thus, we need a mechanism that rewards each miner \emph{after} consensus has been reached on the DAG.
One possible solution is to build a script in each block which will be able to calculate the exact reward based on the three factors. 
But this solution costs too much storage overhead for a block with only one transaction since a script itself will contain address and public key accumulating to about the similar size as a transaction.
It is also computationally expensive since we find the \texttt{secp256k1 ECDSA} key verification time consuming.
We hence propose the following registration-redemption solution
The first block each miner creates (the one directly points to the genesis) should contain a special transaction called \emph{registration}, which specifies an address where future rewards on the peer chain should be awarded to.
Every once in a while, a miner can choose to create a block on his peer chain contain another special transaction called \emph{redemption}, which plays two roles. 
The first is to redeem the accumulated reward since the previous redemption (registration) transaction on his peer chain to the address specified by the previous redemption (registration) transaction. 
The second is to specify the address to which next redemption shall be awarded to. 
To publish the a redemption block block, a miner has to prove the ownership of the address in his previous redemption block by using his secrete key.
As such, the reward of a miner can only be claim by himself. 
Also, a miner can choose to redeem whenever he wants and have the option to specify a different address to store his reward every time he redeems. 
Since a redemption block can only redeem all the rewards in the blocks on the peer chain since the previous redemption block, this effectively prevent a miner from forking his own peer chain since in that way, the forked block costs computing power but brings no reward. 

This design also prevents the forking attack on a peer chain by another miner.
Suppose a malicious miner, Alice, try to fork the peer chain of an honest miner Bob. 
Alice first spends some hashing power to create a block whose \peer{} is set to Bob and pointer $\idp$ is set to point to some block on Bob's peer chain.
If the chosen block is not the head of Bob's peer chain, Alice creates a fork on Bob's peer chain. Alice can surely extend this fork by continuing spending mining power to create more blocks along it. 
Note that Alice cannot publish a redemption block on Bob's chain since that requires Bob's private key to prove the ownership of the address in Bob's previous redemption block.
Thus there is no way Alice can redeem any reward on Bob's chain, even the reward in the forked blocks which are create by Alice.
In fact, this actually gives Bob a choice if the rewards along the fork is larger, Bob can easily append a redemption block to that fork to redeem the rewards. Of course, Bob has to give up the blocks mined by himself by doing so. 
Such an attack causes no damage to consensus and Bob's revenue and Alice's hashing power will be wasted.

In summary, the registration and redemption approach help to determine a unique peer chain for each miner by embedding a sequence of redemptions block which requires that miner's signature. 
Note that a simple idea is to a signature in every block in order to determine a unique peer chain. But a signature require substantial space in the block, causing too much storage and computation (for signature verification) overhead. 

\paragraph{Depth-first Search.}
The final preparation before constructing the ledger is to provide an ordering of all the transactions (equivalently blocks). 
Recall in Section~\ref{sec:dag} that blocks are organized in level sets $\lev(1)$, $\lev(2)$, $\lev(3)$, $\ldots$ along the longest milestone chain. So we order the level sets using the heights of their milestones from low to high.
According to the definition \eqref{eq:level-set-ms}, level set $\lev(n)$ is a directed binary tree with the root being the $n$th milestone $\blk_{m,n}$.
Each directed edge in the tree is essentially a pointer in $\{\idp, \idt\}$.
Note that we remove $\idm$ since it either points to a previous level set or is redundant.
We use the depth-first search (DFS) to traverse this tree starting from the root $\blk_{m,n}$ along the directed edges in the order of first $\idp$ and then $\idt$.
Blocks in each level set are ordered according the post-order DFS\footnote{\url{https://en.wikipedia.org/wiki/Tree_traversal}}. 

Assume we obtain an ordered list of transactions following the above level set ordering and DFS ordering within each level set,
\begin{equation*}
  \tx_1,\tx_2,\tx_3, \ldots
\end{equation*}
Let $\ledger_k$ be the ledger constructed based on the first $k$ transactions and $\utxo_k$ denote the set of UTXO based on $\ledger_k$.
We start with $\ledger_1=\{\tx_1\}$ and $\utxo_1$ being the outputs of $\tx_1$.
In order to obtain $(\ledger_{k+1},\utxo_{k+1})$ from $(\ledger_{k},\utxo_{k})$ and $\tx_{k+1}$, we need to perform the following check. If the inputs of $\tx_{k+1}$ are all from $\utxo_{k}$ and $\tx_{k+1}$ passes signature verification, then
\begin{equation*}
  \begin{split}
    \ledger_{k+1}&=\ledger_{k+1}\cup(\tx_{k+1}),\\
    \utxo_{k+1}&=\utxo_{k}
    \cup \{\text{outputs from }\tx_{k+1}\}
    \setminus\{\text{inputs from }\tx_{k+1}\},
  \end{split}
\end{equation*}
otherwise $(\ledger_{k+1},\utxo_{k+1})=(\ledger_{k},\utxo_{k})$.

We would like to point out that any graph traversal algorithm leading to a unique ordering will be good enough for consensus purpose but an advantage of the DFS ordering is that the chronological order of blocks on a peer chain is kept.

\section{Performance Analysis}
\label{sec:analysis}

We now provide a modeling approach to analyze the performance of such a distributed system.
We focus on the three most important measures---consensus, latency and TPS.
Throughout our analysis, we make the following synchronization assumption. 

\paragraph{Broadcast Delay Assumption.} When a peer broadcasts a message of size $\nu$ size to a peer-to-peer network with $n$ peers, $F(t)$ fraction of the peers will receive the message after $t$ amount of time. Clearly, $F(t)$ increases with time $t$. We further assume there exists a finite time $t_0$ such that $F(t_0)=1$. In other words, all peers will be able to receive the message within $t_0$ amount of time.

Note that the broadcast curve $F$ and the bound $t_0$ depend on both the message size $\nu$ and the number of peers $n$. 
We point out that the block size in our DAG (typically less than one kilobyte in our implementation) is much smaller than that of Bitcoin for example, which is one megabyte. 
So our $t_0$ should be quite small. 
By classical results in regular graphs, a block will be able to reach all $n$ honest peers in $O(\ln(n))$ relays in expectation, assuming it is syntactically valid such that all honest peers will relay the block immediately after receiving it.

\subsection{Reaching Consensus}
\label{sec:concensus}

Since all blocks will be confirmed by some milestone block along the longest milestone chain, our level sets essentially play the role of the blocks in Bitcoin. 
So the consensus of our DAG system essentially boils down to that of the block chain.
This has been formally proven by \cite{garay2015bitcoin} who proposed the Bitcoin Backbone Protocol model. 
We will not repeat the proof here.
Instead, we make the connection to convince readers that it is legitimate for us to directly borrow their result with the following assumptions. 

The analysis in \cite{garay2015bitcoin} is based on discrete rounds, which is not too different from our model. 
We can discretize our time line into intervals of length $t_0$ and let time interval $(rt_0,(r+1)t_0]$ be round $r$.
Based on our broadcast delay assumption, blocks produced in round $r$ will reach all honest peers by the end of round $r+1$. 
In the round-based model, we assume all actions are taken at the end of each round. 

Let $f$ be the probability that there exists an honest peer producing one milestone in a round.
The \emph{honest majority} assumption requires that the number of malicious peers $\kappa$, constitute only a small fraction of the total population. 
Specifically, the number of malicious miner
\begin{equation*}
  \kappa \le(1-2f)n
\end{equation*} 
Note that a large $f$ intuitively implies a high chance that honest miners will create more than one milestones miners in the same round. This would increase the chance of their forking the milestone chain due to broadcast delay. 
Thus the larger the $f$, the smaller the proportion of malicious miners required for the following result to hold true. 
In fact, more subtle relationships exist between $f$ and $(n,\kappa)$. Interested readers can refer to the proofs in \cite{garay2015bitcoin} for more details. The good news is that under the honest majority assumption, we have the following result on consensus.
The following property follows from Theorem 15 in \cite{garay2015bitcoin}.

\paragraph{Common Prefix.}
Consider any pair of honest peers $p_1$ and $p_2$ following our protocol to maintain their local DAGs.
Let $\sdag_{p_1}$ be the local DAG of peer~$p_1$ at round $r_1$, and $\sdag_{p_2}$ be the local DAG of peer~$p_2$ at round $r_2$. 
Suppose $r_1\le r_2$ and the height of $\sdag_{p_1}$ is $h$.
For any positive integer $h'$, let $\blk_{m,h'}$ be the milestone of height $h'$ in $\sdag_{p_1}$, and $\confirm_{p_1}(\blk_{m,h'})$ be the DAG confirmed by $\blk_{m,h'}$ in $p_1$'s local DAG. 
There exists a $k$ such that $\confirm_{p_1}(\blk_{m,h-k})\subseteq \sdag_{p_2}$ with probability higher than $1-e^{-\Omega(k/f)}$, where $\Omega(x)$ dominates $x$ as $x$ grows. 

Note that the Bitcoin Backbone Protocol model is well designed for proving the consensus property, as we are only interested in the existence of the number $k$ and the probability $1-e^{-\Omega(k/f)}$ in the above. 
However, it is not suitable for parameter selection.
For example, to ensure the probability $1-e^{-\Omega(k/f)}$ exceeds 99.9\%, we may end up with a too big a $k$ for practical use. 
This is because the estimations in the backbone model are quite conservative. 
We propose a more accurate model in the next subsection for setting a practical parameter $k$.

\subsection{Wasted Capacity}
\label{sec:throughput}
%
Due to synchronization issues, a transaction may be processed by multiple peers and put in several blocks. 
The extra copies are a waste of hashing power.  
This subsection is devoted to giving an upper bound for the wasted capacity under our transaction assignment protocol \eqref{eq:tx_distance}.

In the subsequent analysis, we assume there are $n$ honest miners with equal hashing power.
We also assume that with the current total hashing power, new blocks are created by following a Poisson process with rate $n\mu$ (blocks/unit of time), where $\mu$ denotes the average rate of block creation for each miner. 

Suppose transaction $\tx$ reaches miner~$i$ at time 0.
However, $\tx$ has just been mined by some miner, who has subsequently broadcasted his block $\blk_{\texttt{tx}}$ containing this transaction to the network at time 0.
Suppose it takes $T_i$ amount of time for miner~$i$ to receive this block.
According to our broadcast delay assumption, $T_i$ is a random variable following distribution $F$.
Let us assume the worst case where the transaction fee of \tx{} is so attractive that miner~$i$ will surely mine this transaction whenever he can before time $T_i$.
Let $N_i(t)$ denote the number of blocks miner~$i$ can create during time $(0,t]$. 
Clearly $N_i(\cdot)$ is a Poisson process with rate $\mu$.
During the time $(0,t]$, miner~$i$ will have changed his miner chain head $N_i(t)$ times, which follows the Poisson distribution with rate $\mu t$.
So conditional on $N_i(t)=k$, the probability that miner~$i$ is eligible to work on transaction \tx{} some time during the interval $(0,t]$ is
\begin{equation*}
  \prob(I_i(t)=1|N_i(t)=k)
  =1-(1-\frac{c}{n})^k
  \le
  1-e^{-\frac{ck}{n}},
\end{equation*}
where $I_i(t)$ indicates whether or not miner~$i$ is eligible to work on \tx{}.

Let $A_i$ denote the event where miner~$i$ successfully mines \tx \  before he receives the block by time $T_i$.
We have the conditional probabilities
\begin{align*}
  & \prob(A_i|I_i(T_i)=1,T_i=t) \le 1-e^{-\mu t},\\
  & \prob(A_i|I_i(T_i)=0,T_i=t) = 0
\end{align*}
where $1-e^{-\mu t}$ is the probability that miner $i$ will successfully create the block in time $t$ and this is an upper bound for the left side of the first expression.

So, conditional on $T_i=t$, the probability that miner~$i$ creates a block for transaction \tx{} can be bounded as follows:
\begin{align*}
	\prob(A_i|T_i=t) &\le (1-e^{-\mu t})\prob(I_i(T_i)=1|T_i=t) \\
	&= (1-e^{-\mu t})\Ex_{N_i}[1-(1-\frac{c}{n})^{N_i}|T_i=t] \\
	&= (1-e^{-\mu t})(1-e^{-\mu t\frac{c}{n}})
\end{align*}
This implies
\begin{equation*}
  \prob(A_i)
  \le \Ex(1-e^{-\mu T_i})(1-e^{-\frac{c}{n}\mu T_i})
  \le (1-e^{-\mu \bar t})(1-e^{-\mu \bar t\frac{c}{n}}),
\end{equation*}
where $\bar t=\Ex(T_i)=\int_0^{t_0} tdF(t)$, and the last inequality is obtained by Jenson's inequality as the above conditional probability is convex in $t$.
The probability that $\tx$ is mined exactly once is lower-bounded by 
\begin{equation*}
	\prod_{i=1}^{n} (1-\prob(A_i))\ge (e^{-\frac{c}{n}\mu \bar t}(1-e^{-\mu \bar t})+e^{-\mu \bar t})^n\rightarrow e^{-c\mu\bar t(1-e\mu \bar t)}
\end{equation*}
as $n\rightarrow \infty$. The expected number of copies of mined $\tx$ is upper-bounded by
\begin{equation*}
1+	\sum_{i=1}^{n} \prob(A_i) \le 1+(1-e^{-\mu \bar t})n(1-e^{-\mu \bar t \frac{c}{n}})\rightarrow 1+(1-e^{-\mu \bar t})\mu c \bar t
\end{equation*}
as $n\rightarrow \infty$.
So the proportion of capacity that is wasted is upper-bounded by
\begin{equation}
  \label{eq:TPS}
  \theta(c) = \frac{(1-e^{-\mu \bar t})\mu c \bar t}{1+(1-e^{-\mu \bar t})\mu c\bar t}.
\end{equation}
$c$ is a design parameter, and the smaller $c$ is, the less capacity that is wasted.

\subsection{Latency}
\label{sec:latency}

After a transaction enters the mempool, it will go through three phases before it is finally confirmed in the public ledger. 
Firstly, the transaction has to wait in the mempool until some miner creates a block $\blk$ to store it. We call this waiting time the \emph{queueing latency} $W_1$. 
Next, this block needs to be confirmed by a milestone $\blk_m$. Recall the definition of confirmation given in \eqref{eq:confirm}. We call this period the \emph{infection latency} $W_2$ since it will be analyzed through an infection model. 
Lastly, the milestone $\blk_m$ needs to be extended by a chain of future milestones with a certain number to ensure a certain level of security. We call the last stage \emph{secure latency} $W_3$.

\paragraph{Queueing Latency.}
Suppose new transactions arrive at the mempool following a counting process with a constant rate $\lambda$.
The mempool is essentially a queueing system with arrival rate $\lambda$ and effective processing rate depending on both $n\mu$ and the transaction assignment rule \eqref{eq:tx_distance}.
We now give an estimation of the waiting time in queue $W_1$ based on the idea of the fluid model in queueing theory.

Denote by $Q$ the stable queue length, i.e.\ number of transactions in the mempool.
Whenever a miner tries to find a transaction from the mempool to work on, he will find that the number of transactions he can process follows the binomial distribution with total number of trials $Q$ and success probability $c/n$, which can be approximated by the Poisson distribution with rate $(Qc/n)$ since $Q$ is large and $c/n$ is small. 
So the proportion of time that a miner has to work on an empty block is the probability that the Poisson random variable equals 0, i.e.\ $e^{-Qc/n}$. 
The rate at which blocks containing transactions are generated is therefore $n\mu(1-e^{-Qc/n})$. As noted in the previous section, only $(1-\theta(c))$ of the transactions are distinct, so the rate at which the transactions in the mempool are processed is $(1-\theta(c))n\mu(1-e^{-Qc/n})$.
For the system to have a stable $Q$, it is required that
\begin{equation*}
	n\mu(1-e^{-Qc/n})(1-\theta(c)) = \lambda.
\end{equation*}
Solving the above equation yields $Q = \frac{n}{c} \ln(\frac{n\mu}{n\mu-\lambda/(1-\theta(c))})$. 
By Little's law, the average waiting time of a transaction is 
\begin{equation}
  \label{eq:W1}
  W_1 = \frac{n}{c\lambda} \ln\left(\frac{n\mu}{n\mu-\frac{\lambda}{1-\theta(c)}}\right) 
  = \frac{1}{c}\frac{1}{\rho\mu}\ln\left(\frac{1}{1-\frac{\rho}{1-\theta(c)}}\right),
\end{equation}
where $\rho=\frac{\lambda}{n\mu}$ denotes the traffic intensity. 
Note that the number of miners does not affect the queueing latency as long as the rate at which new blocks are produced remains the same.
The influence of $c$ on $W_1$ is complicated. On the one hand, a larger $c$ will result in less idle time, thus increasing the effective processing rate. On the other hand, a larger $c$ leads to a higher proportion of duplicate blocks, thus decreasing the effective processing rate. The relation between wasted capacity and queueing latency can be described as follows:
\begin{equation}
	W_1(\theta) = \frac{(1-\theta) \bar t (1-e^{-\mu\bar t})}{\theta\rho}\ln\left(\frac{1}{1-\frac{\rho}{1-\theta}}\right)
	\label{trade-off}
\end{equation}
Our quantitative modeling analysis sheds the light on how the parameter $c$ can be chosen to strike a balance between collision and latency.

Traditional queueing theory suggests that the waiting time will blow up when the traffic intensity $\frac{\rho}{1-\theta}$ approaches 1. A more complicated model can be analyzed by allowing each transaction to have an expiration clock, without which the mempool size will grow without a bound. 
For the time being, the above queueing model is good enough to get the system started. 


\paragraph{Infection Latency.}
A primary difference between the DAG and chain structures is that the former goes beyond the one-dimensional linear structure by allowing parallelism.
Despite the many advantages of the DAG, one issue is that a block may not necessarily be confirmed by the next milestone---it may have to wait for a later milestone. 
A natural question is how long it takes for a block to be confirmed by a milestone block after it is broadcasted to all miners. 
We now provide an upper bound on this waiting time by modeling confirmation in our DAG using an infection model.
In our analysis, we assume all of the $n$ miners are incentivized to follow the three principles specified in Section~\ref{sec:DAG-to-Ledger}.

Suppose that at time $0$, block $\blk$ is in the pending set, i.e.\ it has not been confirmed by any milestone.
As previously discussed, new blocks arrive following a Poisson process with rate $n\mu$.
Each new block has probability $p$ of being a milestone.
Suppose that at time $s>0$, there are $X_s$ miners whose head block can reach $\blk$ by following a path in the DAG.
We say that these miners are \emph{infected} by $\blk$.
Note that if all miners are infected, then $\blk$ will surely be confirmed by the next milestone. 
Assume the next new block is created by miner~$a$ at time $t>s$. With probability $\frac{n-X_{s}}{n}$, miner~$a$ is not infected. 
In this case, the probability that he will become infected assuming he randomly picks any of the $n$ chain heads is $\frac{X_{s}}{n}$.
Note that this probability is lower than the actual probability because block $\blk$ may have already been confirmed by a milestone if the chain head of any of the $X_s$ infected miners's is a milestone. 
Since we are considering an upper bound, we may assume $X_0=1$ and 
\begin{equation}
  X_r = 
  \begin{cases}
    X_s+1\ & \textrm{with probability } \frac{X_s(n-X_s)}{n^2},\\
    X_s\ & \textrm{with probability } 1- \frac{X_s(n-X_s)}{n^2}.
  \end{cases}
\end{equation} 
We also introduce $M_s$, which indicates whether or not $\blk$ is confirmed by a milestone. So $M_0=0$ and
\begin{equation}
  \label{eq:M_s}
  M_t = 
    \begin{cases}
      1\ &\textrm{with probability } p\frac{X_s}{n},\\
      0\ &\textrm{with probability } 1-p\frac{X_s}{n}.
    \end{cases}
\end{equation}
The above modeling gives a continuous-time Markov chain $(X_t, M_t)$ and we are interested in the expected time taken to hit the set $\left\{ (x,m):x,\ge 1, m=1 \right\}$. 
Let $q_x$ denote the expected jumps needed to hit the set starting from $(x,0)$. Then
\begin{align*}
  q_x &= 1+\left(1-p\frac{x}{n}\right)
  \left(
    \frac{x(n-x)}{n^2} q_{x+1}+(1-\frac{x(n-x)}{n^2})q_x
  \right), 
  \quad\  x=1,2,\cdots, n-1, \\
  q_n &= \frac{1}{p}.
\end{align*}
Therefore, 
\begin{align*}
  q_1 &= \sum_{k=1}^n\frac{n^3}{pk^3-n(p+1)k^2+n^2(p+1)k}
         \prod_{j=1}^{k-1}\frac{(pj-n)(j-n)}{pj^2-n(p+1)j+n^2(p+1)}\\
      &< 2n(1+ \ln(n))+\frac{1}{p}.
\end{align*}
Let $\tau_1 := \inf\{t:M_t=1\}$, the time needed for $\blk$ to be confirmed by a milestone. Since blocks arrive at the rate $n\mu$, which is exactly the rate for all jumps, the infection latency is
\begin{equation}
  \label{eq:W2}
  W_2 = \Ex(\tau)
      = \frac{1}{n\mu}q_1 
      < \frac{2+2\ln(n)}{\mu}+\frac{1}{np\mu}.
\end{equation}
Note that the second term in the above upper bound is basically the expected time it takes for a milestone to arrive, which is fixed. 
Another contribution to the infection latency comes from $\ln(n)$ in the first term. Intuitively, the infection latency increases with the number of miners, $n$, due to parallelism. The relationship is better than linear in that it is a slow logarithmic increase.

\paragraph{Secure Latency.}
After a block is confirmed by a milestone, we still need to wait for this milestone to be extended by a number of future milestones for security guarantee.
This is essentially the same latency that occurs in blockchain systems like Bitcoin.
This type of latency was analyzed in \cite{nakamoto2008bitcoin} in a simple model assuming honest miners will not fork among themselves.
However, honest miners may fork among themselves due to broadcast delay because when an honest miner creates a new block, he may not be aware of a recent block created by another honest miner.
A round-based model was formulated by  \cite{garay2015bitcoin} to handle this situation.
The synchronization assumption is that whatever happened in the previous round will be made known to all honest miners in the present round so that they can act accordingly.
However, such a model requires a worst-case scenario analysis and thus is too conservative for parameter selection.
We now describe a continuous-time model to incorporate the broadcast delay function $F$ and the potential forking among honest miners.

\begin{figure}[htbp!]
  \begin{center}
    \begin{tikzpicture}
      \node[draw,rectangle,minimum size=0.32cm] (a4) at (0,1){*};
      \node[draw,rectangle,minimum size=0.32cm] (b4) at (2,1){*};
      \node[draw,rectangle, minimum size=0.32cm] (c4) at (4,1) {1};
      \node[draw,rectangle,minimum size=0.32cm] (d4) at (6,1){*};
      \node[draw,rectangle,minimum size=0.32cm] (a3) at (0,2) {1};
      \node[draw,rectangle, minimum size=0.32cm] (b3) at (2,2){*};
      \node[draw,rectangle,minimum size=0.32cm] (c3) at (4,2){*};
      \node[draw,rectangle,minimum size=0.32cm] (d3) at (6,2){1};
      \node[draw,rectangle,minimum size=0.32cm] (e3) at (8,2){1};
      \node[draw,rectangle,minimum size=0.32cm] (f3) at (10,2){1};
      \node[draw,rectangle,minimum size=0.32cm] (a2) at (0,3) {*};
      \node[draw,rectangle,minimum size=0.32cm] (b2) at (2,3){*};
      \node[draw,rectangle,minimum size=0.32cm] (o) at (-2,2) {1};

      \draw [color=black,-,dashed](-1,0)-- (-1,3.5);
      \draw [color=black,-,dashed](1,0)-- (1,3.5);
      \draw [color=black,-,dashed](3,0)-- (3,3.5);
      \draw [color=black,-,dashed](5,0)-- (5,3.5);
      \draw [color=black,-,dashed](7,0)-- (7,3.5);
      \draw [color=black,-,dashed](9,0)-- (9,3.5);
      \draw [color=black,>=stealth,->](b4)-- (a4);
      \draw [color=black,>=stealth,->](c4)-- (b4);
      \draw [color=black,>=stealth,->](d4)-- (c4);
      \draw [color=black,>=stealth,->](b3)-- (a3);
      \draw [color=black,>=stealth,->](c3)-- (b3);
      \draw [color=black,>=stealth,->](d3)-- (c3);
      \draw [color=black,>=stealth,->](e3)-- (d3);
      \draw [color=black,>=stealth,->](f3)-- (e3);
      \draw [color=black,>=stealth,->](b2)-- (a2);
      \draw [color=black,>=stealth,->, dashed](o)-- (-4,2);
      \draw [color=black,>=stealth,->](a4)-- (o);
      \draw [color=black,>=stealth,->](a3)-- (o);
      \draw [color=black,>=stealth,->] (a2)-- (o);
      
      \node [font=\fontsize{8}{8}\selectfont] at (-3.6,0) {height:};
      \node [font=\fontsize{8}{8}\selectfont] at (-2,0) {$k$};
      \node [font=\fontsize{8}{8}\selectfont] at (0,0) {$k+1$};     
      \node [font=\fontsize{8}{8}\selectfont] at (2,0) {$k+2$};
      \node [font=\fontsize{8}{8}\selectfont] at (4,0) {$k+3$};
      \node [font=\fontsize{8}{8}\selectfont] at (6,0) {$k+4$};
      \node [font=\fontsize{8}{8}\selectfont] at (8,0) {$k+5$};
      \node [font=\fontsize{8}{8}\selectfont] at (10,0) {$k+6$};   
    \end{tikzpicture}
    \caption{Illustration of Tags in the Milestone Tree}
    \label{fig:tag}
  \end{center}
\end{figure}
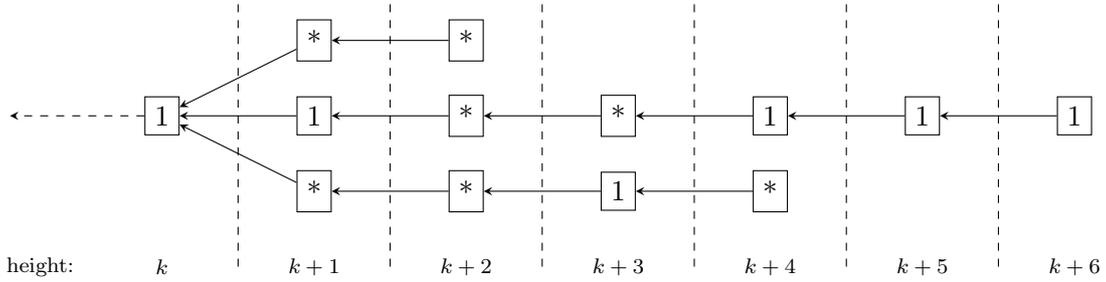

Consider the arrival process of milestones created by honest peers. Let us call such milestones \emph{honest} milestone blocks. 
The creation of honest milestone blocks is a Poisson process with rate $pn\mu$.
Let $U_1, U_2, U_3,\ldots$ denote the inter-arrival times of milestones in this process. 
They are independent and follow the exponential distribution with rate $pn\mu$. 
Let us call the $i$th arrival \emph{milestone $i$}.
This milestone is chronologically the $i$th milestone created by the honest peers, but it may not be the $i$th milestone ever created in the network due to the exsitence of malicious miners.
We would like to tag milestone~$i$ with a value $Y_i\in\{0,1\}$, for $i=1,2,\ldots$, in such a way that if $Y_i = 1$ then the creator of milestone~$i$ will have received all preceding milestones created by all honest miners upon the creation of milestone~$i$.
The tag $Y_i=1$ implies that milestone~$i$ must be higher than any previous milestone tagged with a 1 because the creator of milestone~$i$ is aware of all previous type-1 milestones when he is creating milestone $i$. 
Intuitively, being tagged with a 1 is a good signal and will likely lead to a height increment of the longest chain. 
It follows from our tagging method that, among all milestones of the same height in the milestone tree, at most one can be tagged with a 1 as illustrated in Figure~\ref{fig:tag}, where * represents a milestone of type-0 or a milestone created by a malicious peer. 

The first arrival is tagged with a 1, i.e., $Y_1=1$, as it is the first milestone created by honest peers after the genesis \gen.
Let us now think about then circumstances in which a future honest milestone, say milestone $i$, may be tagged with a 0. 
Consider the case where its preceding honest milestone is tagged with a 1, i.e., $Y_{i-1}=1$. If the inter-arrival time between milestone $(i-1)$ and milestone $i$, $U_i$, exceeds $t_0$, then the honest miner who creates milestone~$i$ will have certainly received all honest milestones. Suppose $U_i=t\in(0,t_0)$, then according to the broadcast delay assumption, the honest miner who creates milestone~$i$ have probability $F(t)$ of having received milestone $(i-1)$ and thus all preceding honest milestones. So milestone~$i$ will be tagged with a 1 with probability $F(t)$. 
In general, let $Z_i$ be a Bernoulli random variable with success probability
\begin{equation*}
  \prob(Z_i=1) = \int_0^{t_0}F(t)pn\mu e^{-pn\mu t}dt + e^{-pn\mu t_0}.
\end{equation*}
Milestone~$i$ will be tagged with a 1 if $Y_{i-1}=1$ and $Z_i=0$. 
Now consider the case where the preceding honest milestone is tagged with a 0, i.e., $Y_{i-1}=0$.
We would like to be conservative by tagging milestone~$i$ with a 0 whenever $U_i<t_0$. In this case, only when the inter-arrival time $U_i$ exceeds $t_0$ can we be sure that the honest miner who creates milestone~$i$ has received all preceding honest milestones and we will then tag milestone~$i$ with a 1. 
Mathematically, 
\begin{equation}
  \label{eq:0-1model}
  Y_i = 
    \begin{cases}
      0,\  \text{if }  (Y_{i-1} = 0 \textrm{ and } U_i\leq t_0)
                       \textrm{ or }
                       (Y_{i-1} = 1\textrm{ and } Z_i =0),\\
      1,\ \text{otherwise.}
	\end{cases}
\end{equation}

The milestone chain evolves as follows. 
First, a number of milestones are tagged with a 1 meaning the longest milestone chain will grow whenever an honest peer produces a milestone. 
When the first type-0 milestone arrives, it may be of the same height as a previous type-1 milestone, and thus it could potentially lead to forks.
Once a type-0 milestone arrives, all newly arriving milestone blocks will be regarded as useless until another milestone tagged with a 1 arrives, the height of which will exceed that of any milestone previously mined by honest peers. 
Beyond that point in time, the regenerative cycle restarts. 
This is a conservative model, because after a typ-0 milestone arrives, the miner of some subsequent milestone with inter-arrival time less than $t_0$ may be informed of all preceding honest milestones if he is lucky enough, but an inter-arrival time greater than $t_0$ will ensure that he will be informed.
%

In each regenerative cycle, the milestones tagged with a 0 can be regarded as wasted. 
Actually, we can consider the wasted milestone as if they were created by malicious miners. 
In other words, the effective hashing power of honest miners should be discounted by the proportion of the blocks tagged with a 1 in a cycle. 
In each cycle, the number of blocks tagged with a 1 follows the geometric distribution with success probability $\int_0^{t_0}(1-F(t))pn\mu e^{-pn\mu t}dt$ and the number of blocks tagged with a 0 is geometric with success probability $e^{-pn\mu t_0}$. 
Thus, the long-run average proportion of milestones tagged with a 1 can be estimated as
\begin{equation*}
  \frac{e^{-pn\mu t_0}}{e^{-pn\mu t_0}
  +\int_0^{t_0}pn\mu(1-F(t))e^{-pn\mu t}dt}.
\end{equation*}
If $pn\mu = 0.1/s$, $t_0 = 2s$ and $F(t)=t-t^2/4$, then the above fraction will be $0.928$. 
So in a network where 10\% of miners are malicious and 90\% are honest, at least $90\times 0.928\%$ of the hashing power would be devoted to growing the chain.
One method to compute the number of milestones we need to wait for is to simply replace $90\%$ with $90\times 0.928\%$ in the Nakamoto model \cite{nakamoto2008bitcoin}.

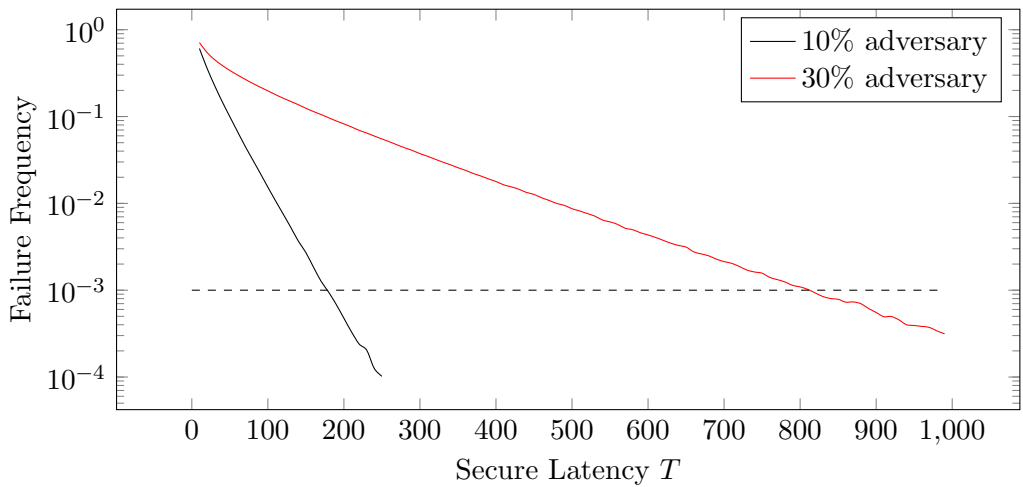
\begin{figure}[H]
  \centering
  \begin{tikzpicture}[scale = 1]
    \begin{semilogyaxis}[
      xlabel= Secure Latency $T$, 
      ylabel = Failure Frequency, 
      x=1cm/100, y=0.5cm]
    \addplot[color=black,
            solid,
            mark=.,
            mark options={solid},
            smooth
            ] coordinates 
            {
(10,0.60783)
(20,0.360404)
(30,0.22717)
(40,0.148161)
(50,0.099533)
(60,0.067301)
(70,0.045709)
(80,0.031853)
(90,0.02221)
(100,0.015359)
(110,0.010738)
(120,0.007566)
(130,0.005317)
(140,0.003687)
(150,0.002713)
(160,0.001841)
(170,0.001267)
(180,0.000947)
(190,0.000683)
(200,0.000475)
(210,0.000331)
(220,0.00024)
(230,0.000204)
(240,0.000125)
(250,0.000102)
            };  
    \addlegendentry{10\% adversary}
    \addplot[
            color=red,
            solid,
            mark=.,
            mark options={solid},
            smooth
            ] 
            coordinates {
            (10,0.71105)
(20,0.546513)
(30,0.451651)
(40,0.388087)
(50,0.33965)
(60,0.301434)
(70,0.268978)
(80,0.241181)
(90,0.217811)
(100,0.197767)
(110,0.178893)
(120,0.162982)
(130,0.149437)
(140,0.137048)
(150,0.124622)
(160,0.114207)
(170,0.105576)
(180,0.096807)
(190,0.088931)
(200,0.082431)
(210,0.075868)
(220,0.069452)
(230,0.06458)
(240,0.059636)
(250,0.055206)
(260,0.051164)
(270,0.047176)
(280,0.04366)
(290,0.040724)
(300,0.037396)
(310,0.034906)
(320,0.032238)
(330,0.029904)
(340,0.027872)
(350,0.025742)
(360,0.024003)
(370,0.022092)
(380,0.020681)
(390,0.01916)
(400,0.017865)
(410,0.016368)
(420,0.01551)
(430,0.014561)
(440,0.013375)
(450,0.01268)
(460,0.011597)
(470,0.010804)
(480,0.00997)
(490,0.009481)
(500,0.008678)
(510,0.008179)
(520,0.007643)
(530,0.007136)
(540,0.006431)
(550,0.00612)
(560,0.005729)
(570,0.005168)
(580,0.004968)
(590,0.004586)
(600,0.004325)
(610,0.004049)
(620,0.003749)
(630,0.003476)
(640,0.003278)
(650,0.003138)
(660,0.002767)
(670,0.002617)
(680,0.002485)
(690,0.002277)
(700,0.002135)
(710,0.00204)
(720,0.001877)
(730,0.001706)
(740,0.001621)
(750,0.00157)
(760,0.001404)
(770,0.001324)
(780,0.001245)
(790,0.001136)
(800,0.001092)
(810,0.001022)
(820,0.000929)
(830,0.000845)
(840,0.000802)
(850,0.000786)
(860,0.000728)
(870,0.000733)
(880,0.000699)
(890,0.000615)
(900,0.000554)
(910,0.000495)
(920,0.000498)
(930,0.000455)
(940,0.0004)
(950,0.000392)
(960,0.000383)
(970,0.000373)
(980,0.000339)
(990,0.000315)
            };    
    \addlegendentry{30\% adversary}
    \addplot[dashed,domain = 0:990] {0.001};
    \end{semilogyaxis}
  \end{tikzpicture}
  \caption{Failure Frequency based on Simulation of $10^6$ Sample Paths}
  \label{fig:simulation}
\end{figure}

Alternatively, we can perform simulations based on our model to determine how long the secure latency needs to be in order to ensure a given level of security.
Suppose a milestone $\blk_m$ is created at time $t$, and is still part of the longest chain in the local DAG of some honest peer after a period of time $T$.
We want $\blk_m$ to be in the longest milestone chain from every honest peer's perspective and to stay in the longest chain thereafter with a high probability. 
Suppose at time $t+T$, milestone $\blk_m$ is on the currently longest chain $C_1$ of one honest peer, and there exists another honest peer adopting a chain $C_2$ which does not contain $\blk_m$.
Consider the type-1 arrivals during $[t+t_0, t+T-t_0]$, all of them are higher than $\blk_m$ and are received by all honest peers by $t+T$. 
Suppose that $\blk$ is one of them and is of height $h>\eta(\blk_m)$. 
Both $C_1$ and $C_2$ have a block of height $h$, say $\blk_1$ and $\blk_2$ respectively, since both peers have received $\blk$ and $C_1$ and $C_2$ are the longest chain from each peer's own point of view.
In addition $\blk_1 \ne \blk_2$ as both of them are higher than $\blk_m$ but only $\blk_1$ extends $\blk_m$.
Hence either $\blk \ne \blk_1$ or $\blk \ne \blk_2$ or both are true.
Therefore the number of 1s during $[t+t_0,t+T-t_0]$ is less than the total number of 0s plus the number of milestones created by malicious miners during $[t, t+T]$.
The probability of this event is low when $T$ is sufficiently large and decays rapidly with respect to $T$ as shown in our simulation results in Figure~\ref{fig:simulation}.
To be on the conservative side, we start our simulation with the first tag being 0, which is stochastically a worse initial condition.
It can be seen from the figure that we have to wait $130$ seconds and $810$ seconds assuming 10\% and 30\% of the hashing power comes from malicious miners, respectively, to ensure a failure frequency of less than $10^{-3}$. 
We use $pn\mu = 0.1/s$, $t_0 = 2s$ and $F(t)=t-t^2/4$ in our simulation. This means once a transaction is first confirmed by a milestone, we need to wait for another 81 and 13 milestones on average before finally accepting the milestone assuming 10\% and 30\% of the hashing power are attributed to malicious miners, respectively.

\paragraph{Discussion on Parameter Selection}
We now summarize the above analysis, and provide a concrete example demonstrating parameter selection and its corresponding performances.
We assume that there are $n=1000$ honest miners in the system, with each miner creating blocks at the rate $\mu$ of 1.2 blocks per second. We also assume that the mempool assignment parameter $c=0.01$ in \eqref{eq:tx_distance}.
So the max TPS of the design is approximately $0.983n\mu=1179.6$ as $1-\theta(c)=0.983$ according to \eqref{eq:TPS}.

The total latency is $W=W_1+W_2+W_3$, where $W_1$ is the queueing latency, $W_2$ is the infection latency and $W_3$ is the secure latency.

By assuming that transactions arrive at the rate $\lambda$ of 1000 per second, the queueing latency $W_1$ is approximately 188 seconds according to \eqref{eq:W1}.
We further assume that the expectation of the inter-arrival time of milestones is $1/pn\mu = 10$ seconds (i.e., $p=1/12000$).
So the infection latency $W_2$ is approximately $23$ seconds according to \eqref{eq:W2}.
If we use the broadcast curve $F(t) = t-t^2/4$ as before, then $t_0=2$ and $\bar t =5/3$.
Suppose also that less than 30\% of hashing power comes from malicious miners.
If we want to ensure that the probability of a successful attack is less than $10^{-3}$, then $W_3$ is $810$ seconds according to the simulation shown in Figure~\ref{fig:simulation}.
So the total latency $W$ is approximately 1021 seconds or just over 17 minutes.


\bibliography{library.bib}

\end{document}